\documentclass[fleqn,usenatbib]{mnras}

\usepackage[T1]{fontenc}

\DeclareRobustCommand{\VAN}[3]{#2}
\let\VANthebibliography\thebibliography
\def\thebibliography{\DeclareRobustCommand{\VAN}[3]{##3}\VANthebibliography}

\usepackage{graphicx}	% Including figure files
\usepackage{amsmath}	% Advanced maths commands
\usepackage{amssymb}	% Extra maths symbols

\usepackage{newtxtext,newtxmath}
\newcommand{\gpm}[3]{$#1^{+#2}_{-#3}$}
\def\gsim{\mathrel{\hbox{\rlap{\lower.55ex \hbox {$\sim$}}
                   \kern-.3em \raise.4ex \hbox{$>$}}}}
\def\lsim{\mathrel{\hbox{\rlap{\lower.55ex \hbox {$\sim$}}
                   \kern-.3em \raise.4ex \hbox{$<$}}}}

\title[SGR\,J1935+2154 multi-frequency]{Multi-frequency observations of SGR\,J1935+2154}

\author[Bailes et. al.]{M.\ Bailes$^{1,2}$, C.\ G.\ Bassa$^{3}$, G.\ Bernardi$^{4,5,6}$,
 S.\ Buchner,$^{6}$
 M.\ Burgay,$^{15}$
 M.\ Caleb,$^{7}$\newauthor
 A.\ J.\ Cooper$^{3,8}$,
 G.\ Desvignes$^{9,10}$,
 P.\ J.\ Groot$^{11,12,13}$,
 I.\ Heywood$^{14,5,6}$,
 F.\ Jankowski$^{7}$,\newauthor
 R.\ Karuppusamy$^9$,
 M.\ Kramer$^{9, 7}$,
 M.\ Malenta$^{7}$,
 G.\ Naldi$^{4}$,
 M.\ Pilia$^{15}$,
 G.\ Pupillo$^{4}$,\newauthor
 K.\ M.\ Rajwade$^{7}$,
 L.\ Spitler$^9$,
 M.\ Surnis$^{7}$,
 B.\ W.\ Stappers$^{7\dagger}$,
 A.\ Addis$^{16}$,
 S.\ Bloemen$^{11}$,\newauthor
 M.\ C.\ Bezuidenhout$^{7}$,
 G.\ Bianchi$^{4}$,
 D.\ J.\ Champion$^9$,
 W.\ Chen$^9$,
 L.\ N.\ Driessen$^{7}$,
 M.\ Geyer$^{6}$,\newauthor
 K.\ Gourdji$^{8}$,
 J.\ W.\ T.\ Hessels$^{3, 8}$,
 V.\ I.\ Kondratiev$^{3, 17}$,
 M.\ Klein-Wolt$^{11}$,
 E.\ K\"ording$^{11}$, \newauthor
 R.\ Le Poole$^{18}$,
 K.\ Liu$^9$,
 M.\ E.\ Lower$^{1,2,19}$,
 A.\ G.\ Lyne$^{7}$,
 A.\ Magro$^{20}$,
 V.\ McBride$^{21}$,\newauthor
 M.\ B.\ Mickaliger$^{7}$,
 V.\ Morello$^{7}$,
 A.\ Parthasarathy$^9$,
 K.\ Paterson$^{22}$,
 B.\ B.\ P.\ Perera$^{23}$,\newauthor
 D.\ L.\ A.\ Pieterse$^{11}$,
Z.\ Pleunis$^{24, 25}$,
A.\ Possenti$^{15,26}$,
A.\ Rowlinson$^{3, 8}$,
M.\ Serylak$^{6}$,\newauthor
G.\ Setti$^{4,27}$,
M.\ Tavani$^{28,29}$,
R.\ A.\ M.\ J.\ Wijers$^{8}$,
 S.\ ter Veen$^{3}$,
 V.\ Venkatraman Krishnan$^9$,\newauthor
 P.\ Vreeswijk$^{11}$,
 P.\ A.\ Woudt$^{12}$\\
$\dagger$ Corresponding Author\\
Affiliations are listed at the end of the paper. 
}

\date{Accepted XXX. Received YYY; in original form ZZZ}

\pubyear{2020}

\begin{document}
\label{firstpage}
\pagerange{\pageref{firstpage}--\pageref{lastpage}}
\maketitle

\begin{abstract}
Magnetars are a promising candidate for the origin of Fast Radio Bursts (FRBs). The detection of an extremely luminous radio burst from the Galactic magnetar SGR~J1935+2154 on 2020 April 28 added credence to this hypothesis. We report on simultaneous and non-simultaneous observing campaigns using the Arecibo, Effelsberg, LOFAR, MeerKAT, MK2 and Northern Cross radio telescopes and the MeerLICHT optical telescope in the days and months after the April 28 event. We did not detect any significant single radio pulses down to fluence limits between 25~mJy~ms and 18~Jy~ms. Some observing epochs overlapped with times when X-ray bursts were detected. Radio images made on four days using the MeerKAT telescope revealed no point-like persistent or transient emission at the location of the magnetar. No transient or persistent optical emission was detected over seven days. Using the multi-colour MeerLICHT images combined with relations between DM, N$_{\textrm{H}}$ and reddening we constrain the distance to SGR~J1935+2154, to be between 1.5 and 6.5~kpc. The upper limit is consistent with some other distance indicators and suggests that the April 28 burst is closer to two orders of magnitude less energetic than the least energetic FRBs. The lack of single-pulse radio detections shows that the single pulses detected over a range of fluences are either rare, or highly clustered, or both. It may also indicate that the magnetar lies somewhere between being radio-quiet and radio-loud in terms of its ability to produce radio emission efficiently.
\end{abstract}


%%%%%%%%%%%%%%%%%%%%%%%%%%%%%%%%%%%%%%%%%%%%%%%%%%

%%%%%%%%%%%%%%%%% BODY OF PAPER %%%%%%%%%%%%%%%%%%

\section{Introduction} 
\label{sec:intro}

Magnetars are possibly the most exotic objects of all the classes of neutron stars. High magnetic fields ($\gtrsim 10^{14}$ G) and intense X-ray/soft $\gamma$-ray emission combined with erratic spin down behaviour makes them unique amongst the neutron star population. The Soft Gamma-ray Repeater (SGR) J1935+2154 was discovered when the \textit{Swift} Burst Alert Telescope detected an X-ray burst on 2014 July 5 \citep{smp+14, lbb+14}. This initial burst was then followed by a number of other short bursts \citep{Cummings+14}. Following this, the sky location of SGR\,J1935+2154 was quickly recognised to place it coincident with the supernova remnant (SNR) G57.2+0.8 strongly suggesting an association \citep{Gaensler+14}. Follow-up radio observations with the Very Large Array (VLA) at 6 GHz revealed no persistent radio emission down to 41 $\mathrm{\upmu Jy}$ \citep[3-$\sigma$;][]{FB+14}. Subsequent observations with the Chandra X-ray telescope detected X-ray pulsations with a period of 3.24\,s \citep{irz+14}. Searches with the Giant Metrewave Radio Telescope (GMRT), Ooty and Parkes telescopes failed to detect any pulsed radio emission \citep{skm+14, bir+14}. Since the initial X-ray outburst the source has undergone further outbursts in 2015 February, 2016 May and 2019 November. It is noteworthy that sporadic X-ray bursts are seen \citep{kis+16, ykj+17} even between these periods of activity, making it one of the most active magnetars known \citep{lgr+20}. As well as the single pulse and periodicity searches, searches for continuum radio emission, that might be associated with a new pulsar wind nebula activated by the bursting activity, were carried out just after the discovery of the source, with the GMRT \citep{sjm+16}. They report 3-$\sigma$ upper limits of 1.2 and 4.5\,mJy at frequencies of 362.5 and 610\,MHz respectively. Searches for continuum emission associated with this most recent burst have been undertaken by a wide range of telescopes and we summarise these in Table~\ref{tab:continuumobs}. In spite of deep searches, no continuum emission has been detected at the best known position of SGR J1935+2154. A detection of pulsar-wind nebula emission from the magnetar would have significant implications on the emission physics of the magnetar and its association with G57.2+0.8 although a pulsar-wind nebula has been detected around only one magnetar to date~\citep{ykk+16}.

SGR\,J1935+2154 became active again from 2020 April 27 when multiple bursts were detected with a number of high-energy telescopes \citep[e.g.][]{hmg+20, pal20,ygk+20}. On 2020 April 28 a very bright serendipitous radio burst, initially thought to be about a kJy\,ms in fluence, was detected by the Canadian Hydrogen Intensity Mapping Experiment (CHIME) telescope in a far sidelobe, but at a location that was consistent with the position of SGR\,J1935+2154 with a dispersion measure (DM) of 332.8~pc~cm$^{-3}$ \citep{scholz20}. The burst was also independently detected by the Survey for Transient Astronomical Radio Emission 2 (STARE2) radio telescope \citep{bkr+20a} with a reported fluence greater than 1.5 MJy\,ms and a very similar DM of 333.2$\pm$0.8~pc~cm$^{-3}$. Subsequently, a targeted observation of SGR\,J1935+2154 by the Five-hundred-metre Aperture Spherical Telescope (FAST) in China resulted in the detection of a fainter $\sim30$~mJy, highly polarised radio burst \citep{zjm+20} indicating that it may be transitioning to a radio loud magnetar state. The bright radio burst detected with CHIME in their 400--800\,MHz band exhibits two components separated by 28.91$\pm$0.02\,ms \citep{abb+20}. They show very different spectral indices with the leading component fading at the highest frequencies, while the trailing component brightens at the highest frequencies. In a subsequent analysis they report average fluences of 480~kJy\,ms and 220\,kJy\,ms for the two bursts respectively with a systematic uncertainty of about a factor of two. The burst detected by STARE2 in the frequency range 1281--1468\,MHz is single peaked and the updated fluence is reported as 1.5(3)~MJy\,ms \citep{brb+20}. These fluences correspond to isotropic equivalent energies of 3$\times10^{34}$\,erg for an assumed distance of 10\,kpc for the CHIME burst \citep{abb+20} and 2.2$\times10^{35}$\,erg for an assumed distance of 9.5\,kpc for the STARE burst \citep{brb+20}. This burst is therefore brighter than any radio burst seen from any Galactic source to date and the corresponding energy is between one and two orders of magnitude less than the equivalent energy for the faintest Fast Radio Bursts (FRBs). This has led to the suggestion that this burst can be linked to the low-end of a FRB luminosity function thereby associating at least some FRBs with magnetars.

The radio burst detected on 2020 April 28 was temporally coincident with the real-time detection of a bright and hard X-ray burst by the INTEGRAL Burst Alert System \citep{msf+20}, and an X-ray burst from a refined analysis of the \textit{Insight}-HXMT\footnote{\url{http://en.hxmt.cn/bursts/331.jhtml}}, AGILE \citep{TUV+20} and Konus-Wind \citep{rsf+20} light curves \citep{zxl+20}. The INTEGRAL X-ray burst light curve in the range 20--200 keV exhibits two peaks separated by $\sim 30$~ms, which is consistent with the separation between the two burst components detected by CHIME \citep{abb+20, msf+20}. Similarly, \textit{Insight}-HXMT reports likely X-ray and hard X-ray counterparts (27--250~keV) to the double-peak radio pulse detected by CHIME. Based on the geocentric arrival times of the CHIME and STARE2 pulses, the two component CHIME burst is well aligned, but not perfectly, with the two peaks in the INTEGRAL light curve while the single peaked STARE2 pulse aligns with the second peak in the INTEGRAL light curve. \textit{Insight}-HXMT was also observing \citep{ltg+20} when FAST detected the highly polarized radio pulse. No significant events were found in the X-ray light curves around the time of the radio detection.

The detection of the radio/X-ray burst led to immediate follow-up campaigns with a variety of radio telescopes and some of these radio observations were coordinated with the X-ray observations. In the immediate aftermath, no radio emission or pulses were detected from the source using the Green Bank Telescope (GBT; \citealp{sjb+20}) or the Deep Space Network (DSN) telescope~\citep{pmp+20}. Apart from the one faint, polarized radio pulse, no other bursts were seen in the subsequent few weeks with FAST~\citep{lzw+20} despite there being 29 SGR bursts detected with the \textit{Fermi} Gamma-ray Burst Monitor during the time window of FAST observations. At the same time, the source continued to show X-ray bursting activity, albeit with a decreasing frequency and mostly weaker bursts.  However some were at least as bright as the one seen concurrently with the radio burst~\citep{bcr+20}. Observations with the Westerbork single 25-m dish RT1 (P-band, 313.49--377.49 MHz), the Onsala 25-m telescope (L-band, 1360--1488 MHz), and the Torun 32-m telescope (C-band, 4550--4806 MHz) took place when the source emitted two X-ray bursts with no simultaneous radio detections \citep{kjs+20}. Later in the same campaign, on 2020 May 24, two radio bursts separated by just 1.4 s, that is much less than the pulse period, were detected at 1.3~GHz with a single antenna of the Westerbork Telescope, with fluences of 112$\pm$22\,Jy\,ms and 24$\pm$5\,Jy\,ms respectively. This was followed by a lack of reported pulsed radio emission for almost four months until October 2020 when the source emitted three radio pulses that were detected by the CHIME telescope~\citep{dg+20} and which all arrived within one pulse period. Prompt follow-up by the FAST telescope resulted in the detection of a number of single pulses from the magnetar and were able to independently derive its spin period~\citep{zb+20}. It is thus apparent that SGR\,J1935+2154 is able to emit radio pulses over a range of pulse energies spanning nearly seven orders of magnitude and is possibly transitioning into a radio loud magnetar.

 Here, we present the results of a long radio follow-up campaign with a number of radio telescopes in order to characterize any radio emission from SGR J1935+2154 in the period after the detection  of the April 2020 bright radio burst. The observations are described in Section~\ref{sec:obs} and the results of the observing campaign are presented in Section~\ref{sec:res}. We discuss possible implications of our findings in Section~\ref{sec:discussions} and draw our conclusions in Section~\ref{sec:conclusion}. 

\begin{table*}
	\centering
	\caption{Continuum observations of SGR\,J1935+2154, including their durations, where available. The central frequency and bandwidth of the observations are indicated by F$_{\mathrm{centre}}$ and BW respectively. }
	\label{tab:continuumobs}
	\begin{tabular}{llccccl}
		\hline
		Telescope & \multicolumn{1}{c}{Date} & Duration & F$_{\mathrm{centre}}$ & BW & Flux Density & References\\
	             &	\multicolumn{1}{c}{(UTC)} & (h) & (MHz) & (MHz) & (mJy) & \\
		\hline
	    EVN   & 2020 May 13 & 5.7 & 1670 & 128 & 0.016 & \cite{nmh+20}\\
		MWA   & 2020 May 3  & 4 & 154 & 31 & 14.4 & \cite{ahl+20}\\
		MWA   & 2020 May 4  & 4 & 185 & 31 & 3.8 & \cite{ahl+20}\\
		uGMRT & 2020 May 27 & 3 & 650 & 200 & 0.090/0.060 & \cite{sjb+20, bmc+20}\\
		uGMRT & 2020 May 28 & 3 & 1360 & 200 & 0.054/0.078 & \cite{sjb+20, bmc+20}\\
		uGMRT & 2020 June 8  & 2 & 400 & 200 & 0.465 & \cite{sjb+20b}\\
		VLA & 2020 Apr 29 & -- & 6000 & 4000 & 0.015 & \cite{rhl20}\\
		VLA & 2020 Apr 29 & -- & 22000 & 8000 & 0.018 & \cite{rhl20}\\
		VLA & 2020 Apr 30 & -- & 6000 & 4000 & 0.007 & \cite{rhl20b}\\
		VLA & 2020 Apr 30 & -- & 22000 & 8000 & 0.024 & \cite{rhl20b}\\
		GMRT & 2014 July 14 & 1.25 & 610 & 33 & 1.2 & \cite{sjm+16}\\
		VLA & 2014 July 6 & 1 & 6000 & 4000 & 0.041 & \cite{FB+14}\\
		\hline
	\end{tabular}
\end{table*}

\section{Observations}
\label{sec:obs}

Simultaneous multi-wavelength observations across a wide spectral range and regular monitoring for further radio emission, either in bursts or periodic emission, would provide valuable insights and new information on the young-magnetar to FRB connection. To this end, a number of observing campaigns were initiated. A multi-telescope campaign was undertaken to monitor SGR\,J1935+2154 on the 11th and 15th of May 2020. The magnetar was observed by the Effelsberg (4 - 8 GHz), MeerKAT (900 - 1700 MHz; tied-array beam and imaging), LOFAR (110 - 190 MHz) and NuSTAR (3 - 79 keV) telecsopes on 11th May 2020, and the MeerKAT (900 - 1700 MHz) and \textit{Insight}-HXMT telescope on 15th May 2020 and with the MeerLICHT optical telescope between May 10 and May 21, 2020. Additionally, simultaneous coordinated observations in the radio and X-ray between Arecibo and \textit{Swift}/XRT and NICER were organised on the 13th and 18th of May 2020. The details of these coordinated observations are summarised in Table \ref{tab:observations} and described below. We also undertook a long-term monitoring campaign at the Northern Cross and MK2 radio telescopes which are summarised in Table \ref{tab:ncMK2obs} and also described below. For each observation, we also checked for radio overlap with any X-ray bursts that were detected by all-sky X-ray burst monitors like {\it Swift} Burst Alert Telescope and \textit{Fermi} Gamma-Ray Burst Monitor~\citep{lgr+20, t20}.

\subsection{Arecibo}

We observed SGR\,J1935+2154 using the Arecibo radio telescope over two epochs on the 13th and 18th of May 2020. For the first epoch, we used the L-Wide receiver (1.15--1.73~GHz). Channelized data with full Stokes polarisation information were recorded using the PUPPI back-end ~\citep[see][and the references therein]{ds+13}. The data were sampled at 10.24 $\upmu$s over a usable bandwidth of 600~MHz. We also performed a 90-second calibrator scan before the observations for polarimetric calibration. For the second epoch, we recorded total intensity data with 40.96~$\upmu$s time resolution over the same usable bandwidth and the same frequency range. The details of the observations are presented in Table~\ref{tab:observations}. The recorded data were then searched offline for bright single pulses from the magnetar. The channels corrupted by strong radio frequency interference (RFI) were masked from further processing. Then, the resulting filterbank data were searched using the \textsc{heimdall} single-pulse detection software\footnote{\url{https://sourceforge.net/projects/heimdall-astro/}}, over a DM range of 300--400~pc~cm$^{-3}$ which covers the DMs of the bursts already presented in the literature. All the resultant candidates from the single pulse search above a signal-to-noise (S/N) of 7 were visually inspected for any signs of broadband astrophysical origin.

\subsection{Effelsberg}
\label{sec:Eff}

At the Effelsberg 100\,m Telescope, SGR\,J1935+2154 was observed 14 times from 2020 April 30 at three different frequency bands (see Table \ref{tab:observations}). Signals from the L-, S- and C-band receivers were converted to pulsar search mode data at 32-bit resolution. The wide band CX-receiver signal was mixed down to two 0--2 GHz bands and digitally processed to generate 4096-channel 8-bit pulsar search mode data with a time resolution of 131.072 $\upmu$s. The data were subsequently searched for any dispersed pulses and periodic pulsed emission. To detect periodic emission, the data were folded using an ephemeris constructed from \cite{ier+2016} and using a DM of 332.8~pc~cm$^{-3}$ \citep{scholz2020}. After cleaning the data for impulsive RFI, dispersed radio bursts were searched for using the \texttt{single\_pulse\_search.py} utility from the \textsc{presto}\footnote{\url{https://github.com/scottransom/presto}} \citep{2011Ransom} software suite, over a DM range of 300--400 pc~cm$^{-3}$. Single pulse candidates with S/N above 7-$\sigma$ were visually inspected.

\subsection{LOFAR}
\label{sec:lofar}

We used LOFAR, the Low Frequency Array \citep{hwg+13,sha+11}, to observe
SGR\,J1935+2154 for 6\,hours, split into 2\,hour observations starting
at 01:49, 04:01 and 06:13 UTC on 2020 May 11. The observation was
configured to obtain simultaneous interferometric and beamforming
products from the COBALT correlator and beamformer \citep{bmn+18}, to
allow low-time, high-spatial resolution imaging of the field of
SGR\,J1935+2154 as well as high-time resolution, low-spatial resolution
beamformed observations of SGR\,J1935+2154. The beamformed data products
used signals from the 23 LOFAR High-Band Antenna (HBA) core stations
to form a tied-array beam pointed at SGR\,J1935+2154. A second tied-array
beam was pointed towards the millisecond pulsar and giant-pulse
emitter, PSR\,B1937+21, offset by $1\fdg3$ degrees from SGR\,J1935+2154
to serve as a test of our analysis pipelines. The results of LOFAR
imaging will be presented elsewhere.
Our data analysis follows that detailed in \citet{cab+20}
and \citet{bhk+20}; we will briefly describe it here. Dual-polarization
complex voltages were recorded for 47\,MHz of bandwidth for
frequencies between 120.2 and 167.1\,MHz. To allow simultaneous
interferometry and beamforming, the bandwidth is somewhat reduced from
the observational set-up used in \citet{bhk+20}, which obtained only
beamforming products.  Using \texttt{cdmt} \citep{bph17}, the complex
voltages were coherently dedispersed to a dispersion measure of
$\mathrm{DM}=332.8$\,pc\,cm$^{-3}$ \citep{scholz20,brb+20}
to create dynamic spectra with 163.84\,$\upmu$s and 24.41\,kHz time
and frequency resolution. Radio frequency interference was removed
with \textsc{rfifind} from \textsc{presto}.

To allow for searches over a wide range of pulse widths, as at this DM we might expect significant scattering at LOFAR frequencies, the dynamic
spectra were down-sampled to 655.36\,$\upmu$s time resolution and
97.65\,kHz frequency resolution (using incoherent dedispersion at
$\mathrm{DM}=332.8$\,pc\,cm$^{-3}$), and incoherently dedispersed in
steps of 0.01\,pc\,cm$^{-3}$ for DMs between 325 and
375\,pc\,cm$^{-3}$ using the \texttt{dedisp} algorithm \citep{bbbf12}. Pulses with widths from the native time resolution up to
50\,ms were searched for using a GPU accelerated version of
\textsc{presto}'s \texttt{single\_pulse\_search.py}. To remain
sensitive to pulses with even greater widths, the analysis was repeated on dynamic spectra downsampled to 20.971\,ms, with incoherent
dedispersion steps of 0.2\,pc\,cm$^{-3}$ and pulse widths up to
3.7\,s. Given the nominal 313\,$\upmu$s scattering width of bursts detected at 1324\,MHz \citep{ksj+20}, at LOFAR frequencies scattering would dominate, and pulse widths of order $\sim2$\,s would be expected.

\subsection{MeerKAT tied-array beam}
\label{sec:MKT_TB}

We recorded the beam-formed output of typically 60 MeerKAT antennas with the (PTUSE) backend \citep{2020Bailes} in both fold and search-mode. The search-mode data have a sampling time of $38.28~\upmu \text{s}$, 8-bit resolution, and 1024 frequency channels across the 856~MHz bandwidth. The folded data have 8-s integrations, 1024 phase bins across the profile, and 1024 frequency channels. Both data sets are stored in PSRFITS format \citep{2004Hotan}. We subjected the data to a variety of searches. First, we visually inspected the folded data for pulsed emission after standard RFI excision. Second, we created single-pulse archive files from the search-mode data using an ephemeris based on \cite{ier+2016} (as in Sec. \ref{sec:Eff}) and visually inspected those. Moreover, we ran single-pulse tools on them that we developed for data analysis of rotating radio transient pulsars (Jankowski et al.\ in prep). Finally, we aggregated the data into 1-min integrations and converted them to \textsc{sigproc}\footnote{\url{https://sigproc.sourceforge.net}} filterbank format using tools from the \textsc{presto} software suite. We then searched them using the \textsc{heimdall} software for trial DMs between zero and $600~\text{pc} \: \text{cm}^{-3}$. We utilised standard RFI excision methods, including a static RFI mask that left us with an effective bandwidth of about 630~MHz. We visually inspected all resulting candidates with S/N above 7, pulse widths up to about 39.2~ms, and DMs between 300 and $360~\text{pc} \: \text{cm}^{-3}$, i.e.\ in a window around the nominal DM of the magnetar.

Additionally, during the observations on UTC 2020 May 8, 11 and 15, the MeerTRAP single-pulse pipeline running on the TUSE instrument (Transients User Supplied Equipment; Stappers et al.\ in prep) undertook a real-time search for single pulses. Using the innermost 40 MeerKAT dishes it searched 768 full-bandwidth (see Table \ref{tab:observations}) beams formed by the FBFUSE backend (Filter-bank Beamformer User Supplied Equipment; Barr et al.\ in prep). These data had a time resolution of $306.24~\upmu \text{s}$. The beams were placed so that they overlapped at 25~per cent of the peak sensitivity and covered a total area of $0.8~\deg^2$ centred on the position of SGR\,J1935+2154.

\begin{figure*}
\centering
\includegraphics[width=\textwidth]{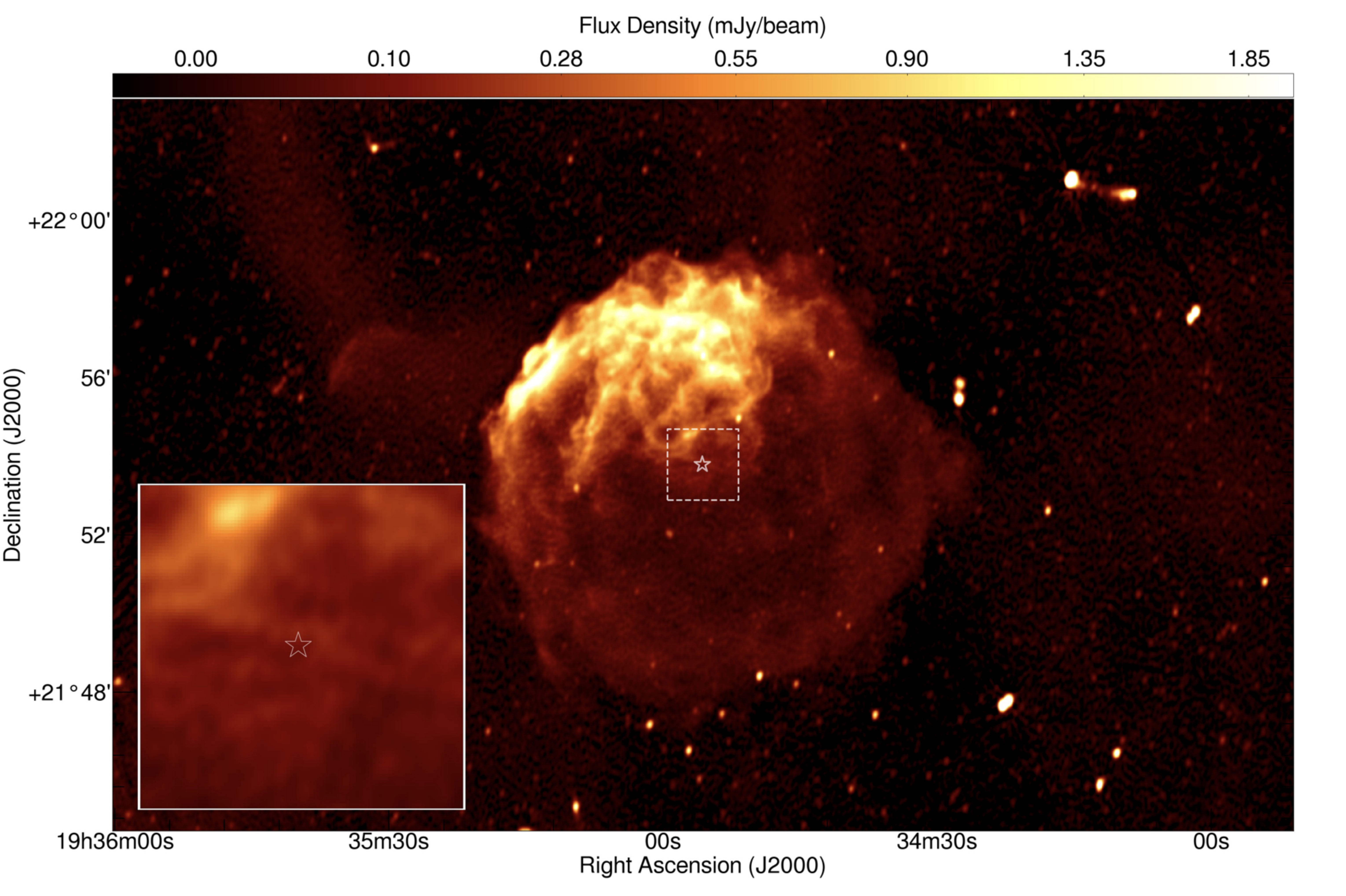}
\caption{The inner 0.5$^{\circ}$~$\times$~0.3$^{\circ}$ of the target field, formed by combining the 11 May 2020 and 15 May 2020 MeerKAT observations. The position of SGR\,J1935+2154 is marked with a star, embedded in the emission from SNR G57.2+00.8 which dominates the field. The thermal noise in this image is approximately 5.2 $\mu$Jy beam$^{-1}$. The restoring beam in this image is a 2D Gaussian with an extent of 8.4$''$~$\times$~5.8$''$ (position angle 178 degrees, east of north). The inset panel shows a 1.8$'$~$\times$~1.8$'$ zoom on the location of SGR\,J1935+2154, as marked by the dashed box on the main panel. The colourmap has a square-root stretch function, and is the same for both panels. Pixel values in mJy beam$^{-1}$ are indicated at the top of the figure.}
\label{fig:meerkat_image}
\end{figure*}

\subsection{MeerKAT imaging}

Imaging-mode observations of SGR\,J1935+2154 were carried out with MeerKAT on three occasions (see Table \ref{tab:observations} for the dates). The data reduction procedure\footnote{MeerKAT calibration and imaging scripts are available here: \url{https://ascl.net/2009.003} \citep{heywood20}.} was consistent for each epoch. The visibilities were averaged from their native 4096 channels to 1024 channels. Basic flagging commands were applied to all the data using the {\sc CASA} package \citep{mcmullin07}, including the flagging of persistent RFI bands on baselines below 600~m. Auto-flagging was performed on the calibrators using the {\sc tfcrop} and {\sc rflag} algorithms within {\sc CASA} using their default settings. Delay and bandpass solutions were iteratively obtained from observations of the primary calibrator PKS~B1934$-$638 with rounds of flagging in between iterations. PKS~B1934$-$638 was also used to set the absolute flux scale. Time-dependent gains were derived from observations of J2011$-$0644, which was observed for 2 minutes for every 20 minutes of target observation. The target data were corrected using these calibration solutions, and then flagged using {\sc tricolour}\footnote{\url{https://github.com/ska-sa/tricolour}}\citep{pmm+20}. 

The target data were imaged using {\sc wsclean} \citep{offringa14} with multiscale cleaning \citep{offringa17} and iterative masking. Phase-only self-calibration solutions were derived for every 128 seconds of data, and amplitude and phase corrections were solved for once per 20 minute scan using {\sc CASA}. These solutions were applied and the imaging process was repeated. To fully model and deconvolve off-axis emission, the images consisted of 10240$^{2}$ pixels of 1.1$\times$1.1 arcsec$^{2}$. A Briggs' weighting parameter of $-$0.8 was used to temper the sensitivity to large scale radio structures. In addition, a Gaussian taper was applied to the ($u$,$v$) plane to obtain a 5 arcsec synthesised beam. No primary beam correction was applied. The inner 0.5$^{\circ}$~$\times$~0.3$^{\circ}$ of the resultant image of the field is shown in Fig. \ref{fig:meerkat_image}.

In addition to the three observations made in `standard' imaging mode, we also obtained visibilities from the beamformer mode observation on 29 April 2020. No calibrator scans were available for this run. The data were simply flagged using {\sc tricolour}, and then the multi-frequency, multi-scale clean component model from the 15 May 2020 run was used to predict model visibilities to self-calibrate the data.

\subsection{MK2 observations}

We also initiated a monitoring campaign of SGR\,J1935+2154 using the 32-m Mark 2 (MK2) radio telescope located at Jodrell Bank Observatory. The source was monitored with almost daily cadence at 1532~MHz in separate 1 hour observations from 2020 May 30, with a consistent observing set-up. For each observation, a 400~MHz band was coarsely channelized into 25 subbands of 16~MHz each with a polyphase filter using a ROACH-based backend~\citep{bjk+16}. For the single pulse search each 16~MHz subband was further channelized into 32 frequency channels using \texttt{digifil} from the \texttt{dspsr} software suite~\citep{sb11}, and downsampled to a sampling time of 256~$\upmu$s. After removing edge channels and known RFI the resulting data span approximately 400~MHz, split into 672 frequency channels. We then masked further frequency channels in the data containing narrow-band RFI. The resulting data were then searched for significant signals (S/N $>$ 7) over a small range of DM (325--335~pc~cm$^{-3}$) around the nominal value for the magnetar using \texttt{heimdall}~\citep{bbb+12}. We also used the digital filterbank (DFB) backend, that incoherently dedispersed and folded the data at the DM and period of the magnetar using the best know ephemeris of SGR\,J1935+2154 from \cite{ier+2016}. After RFI removal, a bandwidth of approximately 336~MHz was used. The operational details of the DFB backend are provided in~\cite{mhb+13}.

\subsection{Northern Cross}

Observations with the Northern Cross (NC) radio telescope were carried out for approximately 1.5~h during transit for 35 days starting on 2020 April 30 and ending on 2020 June 12. The few interruptions during this period (see Table \ref{tab:ncMK2obs}) were due to technical issues or to the use of the antenna for other observations. The system, upgraded to observe FRBs, is described in \citet{locatelli20}. Only a portion of the North-South arm (6 cylinders out of 64) was used in this observing campaign. The system is composed of 24 radio receivers (4 receivers per cylinder) and it has an effective area of $\simeq 750\mathrm{~m}^{2}$. Signals from the 24 receivers were calibrated by observing Cas~A in interferometric mode, and then a single digital beam was formed, that followed the source as it transited across the field of view. Beam-formed voltages were stored to disk with a 138.24~$\upmu$s time resolution. The system operates in the 400--416 MHz frequency band, with 12.2~kHz frequency resolution. The data was searched for bursts as wide as 100~ms using \textsc{heimdall}. Data were also folded in search of pulsations using the spin period measured from the NuSTAR X-ray telescope data \citep{borghese20atel} and the DM of the FRB-like bursts detected on 2020 April 28 \citep{abb+20, brb+20}.

\begin{table*}
	\centering
	\caption{Summary of radio and optical observations. The two Arecibo observations were coordinated with Swift and NICER respectively. The radio burst fluence limits is denoted as RBF assuming a burst width of 1~ms. Both the RBF and the average pulse flux limits (the final column) correspond to a signal-to-noise ratio of 7 (signal-to-noise ratio 5 for the optical observations). Optical limits correspond to the deepest (q) band. See Appendix~\ref{sec:meerlicht} for a full overview of the optical observations.}
	\label{tab:observations}
	\begin{tabular}{lcccccccc} % four columns, alignment for each
		\hline
		Telescope & T$_{\mathrm{start}}$ & T$_{\mathrm{end}}$& F$_{\mathrm{centre}}$ & BW & N$_{\text{chan}}$ & t$_{\mathrm{samp}}$ & RBF & Flux\\
	             &	(UTC) & (UTC) & (MHz) & (MHz) & & ($\upmu \text{s}$) & (mJy~ms) & ($\upmu \text{Jy}$)\\
		\hline
		Arecibo &  2020-05-13 09:13:16 & 2020-05-13 09:56:00 & 1380 & 600 & 4096 & 10.24 & 25& 4.5 \\
		Arecibo &  2020-05-18 08:51:09 & 2020-05-18 09:40:00 & 1380 & 600 & 2048 & 40.96 & 25& 4.3\\
		Effelsberg & 2020-04-30 07:32:40 & 2020-04-30 08:32:20 & 1360 & 240 & 256 & 51.2 & 169.54 & 29.95\\
		Effelsberg & 2020-05-01 00:00:10 & 2020-05-01 00:59:40 & 1360 & 240 & 256 & 51.2 & 169.54 & 29.87\\
		Effelsberg & 2020-05-01 02:39:20 & 2020-05-01 05:51:30 & 1360 & 240 & 256 & 51.2 & 169.54 & 16.64\\
		Effelsberg & 2020-05-06 06:16:10 & 2020-05-06 07:15:20 & 1360 & 240 & 256 & 51.2 & 169.54 & 29.95\\
		Effelsberg & 2020-05-07 05:45:30 & 2020-05-07 09:47:50 & 1360 & 240 & 256 & 51.2 & 169.54 & 14.94\\
		Effelsberg & 2020-05-08 07:02:10 & 2020-05-08 08:02:10 & 4850 & 500 & 512 & 51.2 & 121.94 & 21.29\\
		Effelsberg & 2020-05-11 03:36:54 & 2020-05-11 08:36:10 & 6000 & 4000 & 4096 & 131.072 & 63.63 & 5.00 \\
		Effelsberg & 2020-05-19 01:16:20 & 2020-05-19 02:58:00 & 2550 & 80 &  256 & 102.4 & 200.64 & 27.12\\
		Effelsberg & 2020-05-20 04:55:50 & 2020-05-20 06:55:50 & 1360 & 240 & 512 & 102.4 & 169.54 &21.08\\
		Effelsberg & 2020-05-28 03:31:50 & 2020-05-28 07:31:50 & 2550 & 80 & 256 & 102.4 & 177.34 &22.0\\
		Effelsberg & 2020-06-04 02:41:30 & 2020-06-04 04:41:30 & 1360 & 240 & 256 & 102.4 & 169.54 &21.08\\
		Effelsberg & 2020-06-06 23:18:49 & 2020-06-07 01:18:49 & 2550 & 80 & 256 & 102.4 & 177.34 & 22.05\\
		Effelsberg & 2020-06-08 03:03:39 & 2020-06-08 07:03:49 & 1360 & 240 & 256 & 102.4 & 169.54 & 21.08\\
		Effelsberg & 2020-06-22 23:49:40 & 2020-06-23 04:44:50 & 1360 & 240 & 256 & 102.4 & 169.54 & 12.24\\
		LOFAR      & 2020-05-11 01:49:00 & 2020-05-11 03:49:00 &  144 &  47 & 480 & 655.4 & & \\
		LOFAR      & 2020-05-11 04:01:00 & 2020-05-11 06:01:00 &  144 &  47 & 480 & 655.4 & & \\
		LOFAR      & 2020-05-11 06:13:00 & 2020-05-11 08:13:00 &  144 &  47 & 480 & 655.4 & & \\
		MeerKAT$^\text{a}$ & 2020-04-29 02:39:27 & 2020-04-29 06:58:21 & 1284 & 856 & 1024 & 38.28 & 75 &\\
		MeerKAT$^\text{b}$ & 2020-05-08 05:16:48 & 2020-05-08 06:33:36 & 1284 & 800, 856 & 1024, 4096 & 38.28, 306.24 & 75 & \\
		MeerKAT$^\text{b}$ & 2020-05-11 03:01:01 & 2020-05-11 06:18:32 & 1284 & 800, 856 & 1024, 4096 & 38.28, 306.24 & 75 & \\
		MeerKAT$^\text{b}$ & 2020-05-15 00:41:00 & 2020-05-15 04:00:14 & 1284 & 800, 856 & 1024, 4096 & 38.28, 306.24 & 75 &\\ \hline
		Telescope & T$_{\mathrm{start}}$ & T$_{\mathrm{end}}$& Bands &  & No. Exp.  & t$_{\mathrm{exp}}$ & Lim. Mag. & Lim. Flux\\
	             &	(UTC) & (UTC) & & & & (\text{s}) & (AB) & ($\upmu \text{Jy}$)\\
		\hline
		MeerLICHT &2020-05-10 02:24 & 2020-05-10 03:00 &$u,g,q,r,i,z$ && 3$\times$6& 60 & 19.75 & 45.88\\
		MeerLICHT &2020-05-15 02:49 & 2020-05-15 03:00 &$u,g,q,r,i,z$ && 1$\times$6& 60 & 20.43 & 24.35 \\
        MeerLICHT &2020-05-15 03:11 & 2020-05-15 04:26 & $q$ && 34 & 60 & 20.41 & 24.92 \\
        MeerLICHT &2020-05-17 02:49 & 2020-05-17 03:00 & $u,g,q,r,i,z$ && 1$\times$6 & 60 &20.64 & 20.15 \\
        MeerLICHT &2020-05-17 03:14 & 2020-05-17 04:19 & $q$ && 27 & 60 & 20.73 & 18.60\\
        MeerLICHT &2020-05-18 02:52 & 2020-05-18 03:02 & $u,g,q,i,z$ && 1$\times$5 & 60 & 20.60 & 20.82 \\
        MeerLICHT &2020-05-18 03:05 & 2020-05-18 04:20 & $q$ && 32 & 60 & 20.79 & 17.59 \\
        MeerLICHT &2020-05-19 02:50 & 2020-05-19 03:00 & $u,g,q,r,i,z$ && 1$\times$6 & 60 & 18.40 & 159.06 \\
        MeerLICHT &2020-05-19 03:02 & 2020-05-19 04:19 & $q$ && 25 & 60 &20.70 & 19.14\\
        MeerLICHT &2020-05-21 03:06 & 2020-05-21 03:19 & $u,g,q$ && 1$\times$3 & 60 &20.16 & 31.25\\
		\hline
		\multicolumn{9}{l}{$^\text{a}$ beam-formed}\\
		\multicolumn{9}{l}{$^\text{b}$ imaging \& beam-formed}\\
	\end{tabular}
\end{table*}

\begin{figure*}
  \centering
  \includegraphics[width=\textwidth]{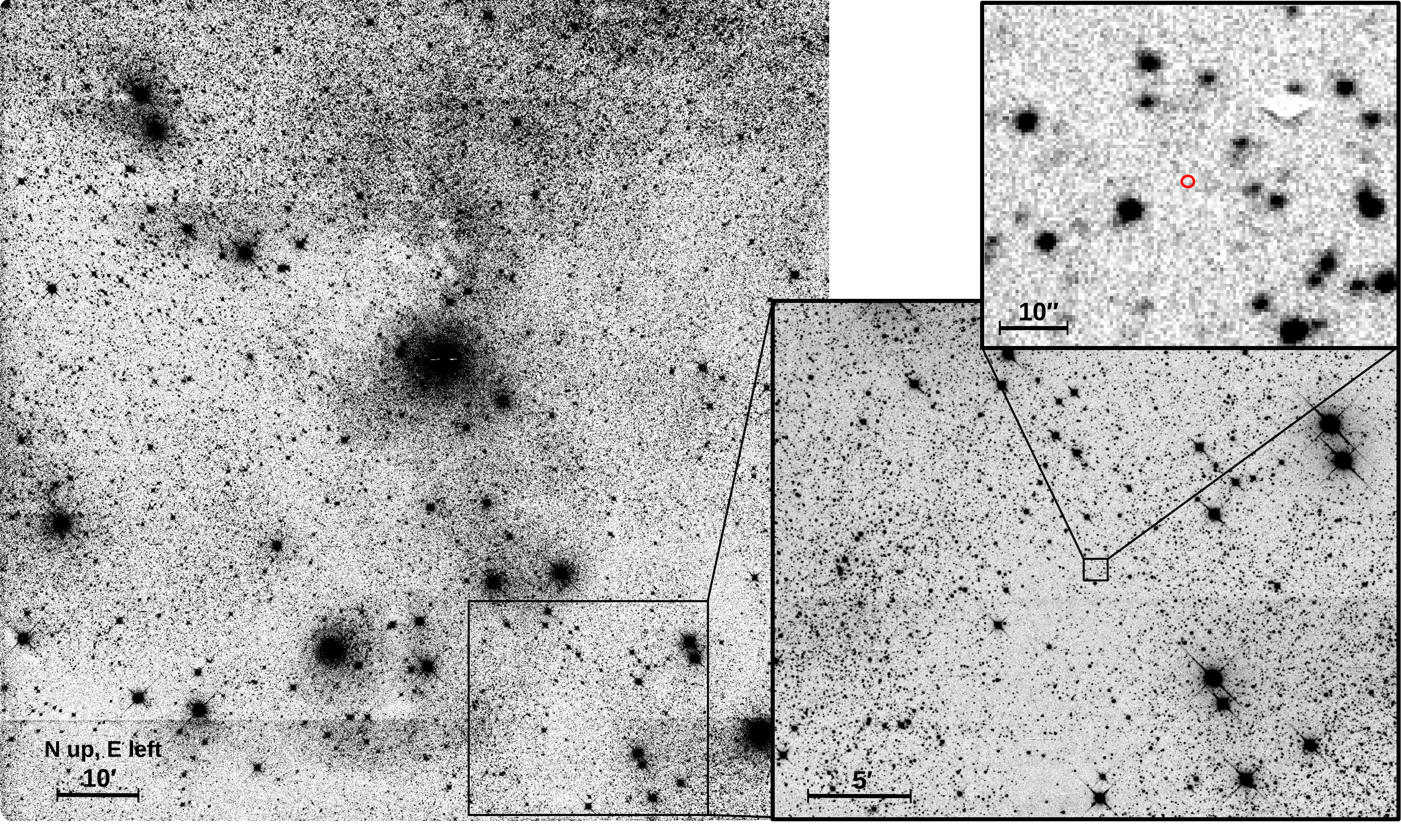}
  \caption{Co-added MeerLICHT $q$-band image of the field of SGR\,J1935+2154. Left, the full field with ID 11744, spanning 1.65$\degr\times$1.65$\degr$. Darker regions, e.g. at the top, are the results of denser star field. Lighter regions show the effect of obscuring dust lanes. The horizontal stripe at the bottom is a low-level artefact of the border between two read-out channels. Bottom right, a 30$^\prime\times$30$^\prime$ zoom-in centered on the position of SGR\,J1935+2154 and matching the size and orientation of the MeerKAT supernova remnant image shown in Fig.~\ref{fig:meerkat_image}. Top right, a 1$^\prime\times$ 1$^\prime$ close-up of the region centered on SGR\,J1935+2154. The red circle indicates the position of SGR\,J1935+2154. Faintest detected sources are at the level of AB$_q$=22.8 magnitude.}
  \label{fig:meerlicht}
\end{figure*}

\subsection{MeerLICHT}  

MeerLICHT is the optical wide-field imager that is twinned with the MeerKAT radio array \citep{bgw+16}. It features a 65\,cm primary mirror and the modified Dall-Kirkham design gives it a field-of-view of 2.7 square degrees, sampled at 0.56"/pix with an STA 1600 10560$\times$10560 CCD. MeerLICHT is equipped with a Sloan-type filter set ($u,g,r,i,z$) and additionally has a wide-band (440--720nm) $q$-band filter. 
Between 2020 May 10 and 21 the field of SGR\,J1935+2154 was observed a total of 162 times in 60s integrations, using all six filters. All data was processed automatically by the {\tt BlackBOX} pipeline, developed for the MeerLICHT and BlackGEM projects. {\tt BlackBOX} performs standard optical calibrations (de-biasing, overscan corrections, flat-fielding) as well as astrometric and photometric calibrations. The astrometry is based on the Gaia DR2 catalog, with after-calibration rms values of <50 mas in both coordinates, across the full field. The photometry is calibrated using a set of 70 million all-sky standard stars that have their spectral-energy distribution reconstructed from Gaia, SDSS, PanStarrs, SkyMapper, Galex and 2MASS catalogs. Uncertainties on the photometric zero-point per image are at the level of 1\%. Output from {\tt BlackBOX} includes a full-source catalog of all detected sources, as well as a transient catalog of transient and/or high-variability sources. The last is based on difference imaging which is performed using the {\tt ZOGY} \citep{zog16} algorithms. As the MeerLICHT/BlackGEM sky grid field containing the position of SGR J1935+2154, with FieldID 11744, had not been observed yet as part of the MeerLICHT Southern All Sky Survey that covers the full sky south of declination +30$\degr$, the first image on the field in each filter is taken to be the reference image for the image differencing routines. If a transient source is also present in the reference image, it is only detected as a transient if it changes it flux significantly between the reference image and the subsequent images. However, it would then also be present in the full-source catalog of objects. {\tt BlackBOX} includes a quality control step, where observations are flagged according to a colour-coding (green, yellow, orange or red) based on a set of numerical estimates on the quality of the observations. For the data obtained here all the $u$-band data were red-flagged due to a too high standard deviation of the photometric zero-point over the field-of-view. This means that all $u$-band data did get processed but no transient detection algorithm was applied. MeerLICHT/BlackGEM magnitudes are on the AB system. Limiting magnitudes are calculated from the detected sources and represent the brightness of a source that would have been detected at the 5-$\sigma$ level. {\tt BlackBOX} calculates both field-level limiting magnitudes, i.e. a single value across the full field, as well as position-dependent limiting magnitudes for transients. For example the position-dependent limiting magnitude can vary significantly from the field-level limiting magnitude for transients located on top of bright galaxies. 

All MeerLICHT observations were analysed on a per-exposure basis. Afterwards co-adds of all the images taken in the same filter were made, for the possible detection of a fainter counterpart. An image of the field from the deep co-add in the $q$-band (440-720nm) surrounding SGR~J1935+21 is shown in Fig\ \ref{fig:meerlicht}. An overview of the MeerLICHT data is shown in Table~\ref{tab:observations}. A detailed log of MeerLICHT observations is given in Table~\ref{tab:meerlicht} in Appendix~\ref{sec:meerlicht}.

\begin{figure*}
  \centering
  \includegraphics[width=\textwidth]{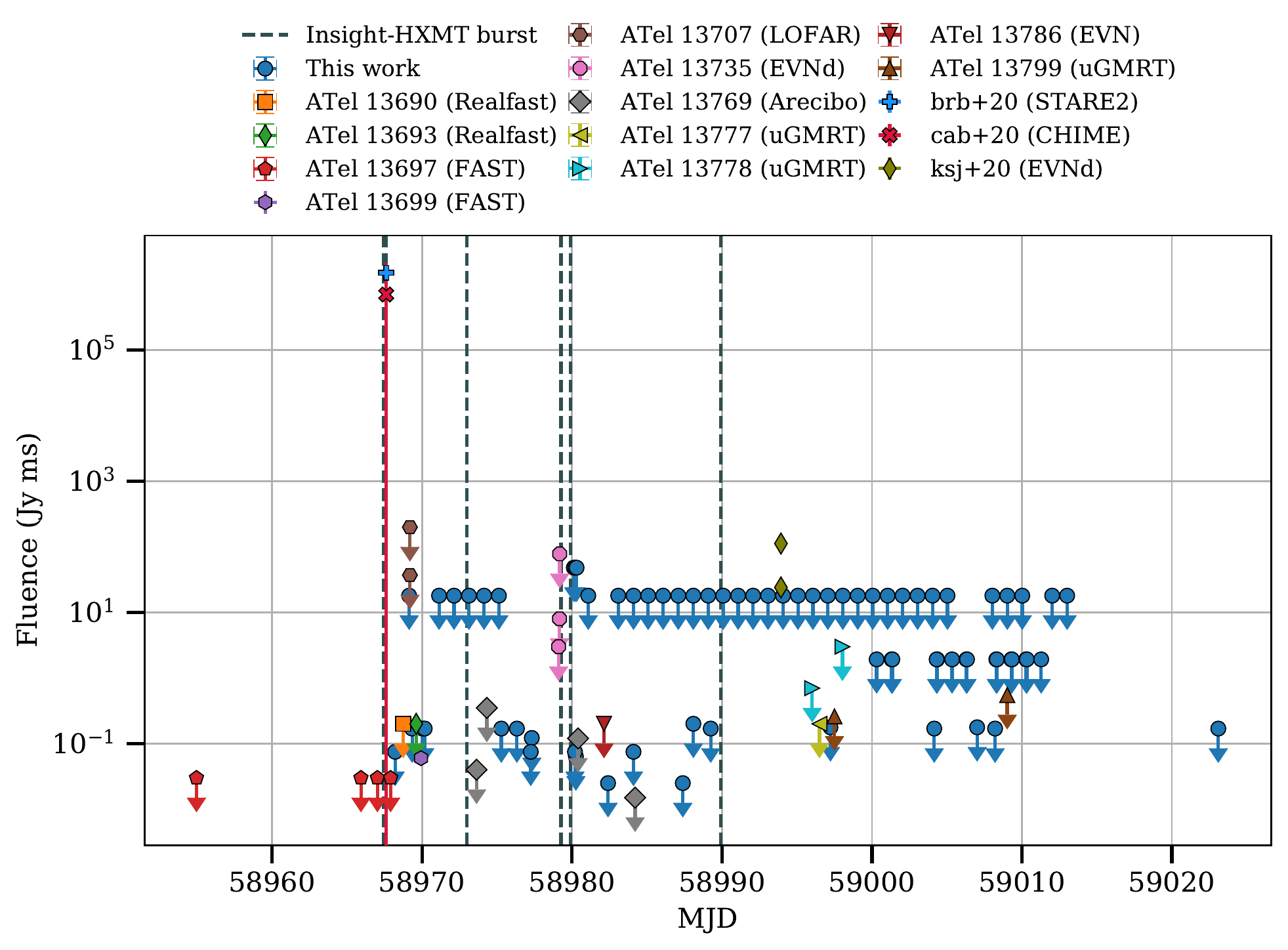}
  \caption{Timeline of the single-pulse fluence upper limits of SGR~J1935+2145 at various radio frequencies as presented in this work, together with a selection of measurements from the literature. Fluences from this work are quoted for assumed burst widths of 1-ms (see Tables~\ref{tab:observations} and \ref{tab:ncMK2obs}), except in the case of the low-frequency LOFAR observations, where pulse scatter-broadening effects likely dominate. Additionally, we mark the times of the brightest X-ray bursts with fluences $> 10^{-7}~\text{erg}~\text{cm}^{-2}$ reported by \textit{Insight}-HXMT.}
  \label{fig:fluencelimits}
\end{figure*}

\section{Results}
\label{sec:res}

\begin{figure}
    \centering
    \includegraphics[width=3.0in]{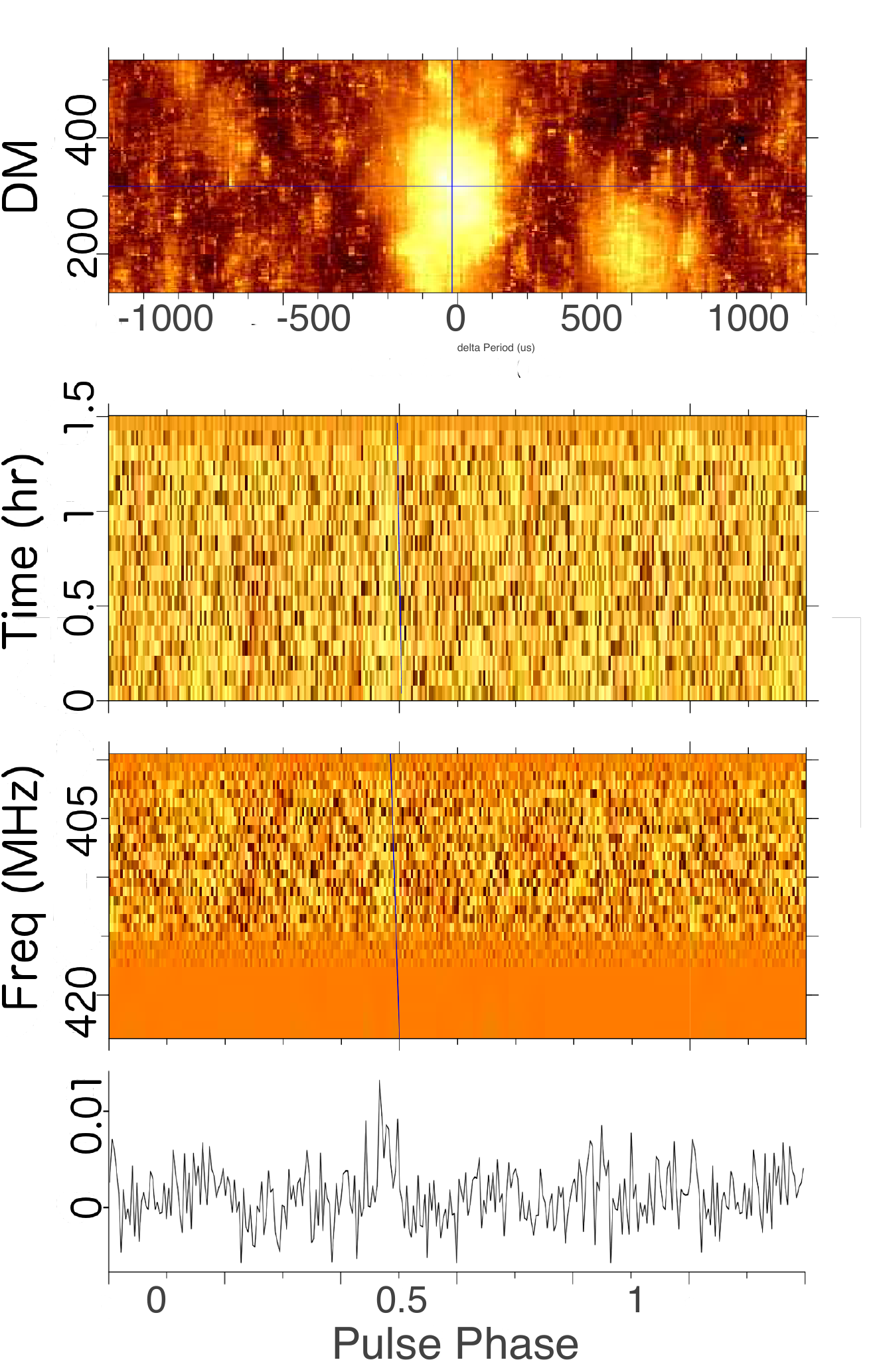}
   \caption{A radio pulse detected during observations of SGR\,J1935+2154 with the  Northern Cross (obtained on 2020 May 30). Bottom to top the plots show: best integrated pulse profile (vertical axis is in arbitrary units); a waterfall plot of the frequency sub-bands versus pulse phase (the lighter the colour, the stronger the signal); a waterfall plot of time sub-integrations (5 minutes each) versus pulse phase; a color scale showing the S/N of the integrated pulse profile in function of the spin period and dispersion measure (brighter colors correspond to higher values). The values of $P$ and DM maximising the S/N and corresponding to the bottom profiles are reported in the text. The optimised value of $P$ is inconsistent with that derived from an ephemeris generated from NICER data.}
    \label{fig:radioDet}
\end{figure}

\subsection{Radio non-detections}

No potential candidates were found in any of the single pulse searches of the Arecibo data from both epochs (see Table \ref{tab:observations}) and we report an upper limit on the fluence of a potential single pulse from the magnetar of 0.09~Jy~ms for a 5 ms wide pulse, under the assumption that the emission spans the available bandwidth. Simultaneous observations took place with the \textit{Swift}/XRT instrument in photon counting mode during the Arecibo time window on 2020 May 11. No coincident bursts were detected \citep{bcr+20}. Similarly, simultaneous observations with the \textit{NICER} X-ray telescope on 2020 May 18 did not result in coincident detections either (A. Borghese, private communication). No periodic or transient signals above 7-$\sigma$ were detected in the Effelsberg observations. The sensitivities of the receivers used for these observations vary across the observing band or between the two orthogonal probes of the receiver\footnote{\url{https://eff100mwiki.mpifr-bonn.mpg.de/doku.php?id=information_for_astronomers:rx_list}}. The detection limits in Table \ref{tab:observations} are based on an average value of 25, 18, 11 and 17 \,Jy for the system equivalent flux density of the CX-, C-, S- and L-band receivers for a signal-to-noise ratio of 7-$\sigma$ and 10\% duty cycle. No bursts were found in either analysis method of the LOFAR observations (see Section \ref{sec:lofar}). We estimate the sensitivity limits for LOFAR by using the radiometer equation and the method detailed in \citet{kvh+16}, which uses the \cite{ham06} model for the LOFAR HBA beam, combined with system temperature estimates, RFI fraction and beamformer coherence. Based on this, we determine 150\,MHz fluence limits of 48\,Jy\,ms for a 20\,ms burst at a S/N of 7, 260\,Jy\,ms for a 200\,ms burst at a S/N ratio of 12, and 440\,Jy\,ms for a 2\,s burst at a S/N of 15. Similarly, no credible pulses were detected in any of our MeerKAT searches, which translates to a peak flux density upper limit of about 75~mJy at a frequency of 1.28~GHz, and corresponds to a fluence upper limit of 75~mJy~ms, assuming 1~ms pulse width. No credible detections were made in the MeerKAT/MeerTRAP real-time searches on 2020 May 8, 11 and 15 either (see Section \ref{sec:MKT_TB} for more details). The analysis of the NC data produced a series of candidates that were visually inspected if they exceeded a S/N of 7. This revealed no detections down to a limiting sensitivity of 18~Jy~ms for a 1~ms burst. For the observations with the MK2 telescope we use a gain of approximately 0.2~K~Jy$^{-1}$, a receiver temperature of 45~K taking into account the excess sky temperature due to the supernova remnant and a duty cycle of 3$\%$. We report a 7-sigma flux density limit of 232.1~$\upmu$Jy for pulsed radio emission. For a 7-$\sigma$ single pulse with a width of 1~ms, we report a peak flux limit of 1.92~Jy. We note that the ephemeris \citep{ier+2016} used for the searches for average pulse emission was somewhat out of date for a variable source like a magnetar, and so all these searches also searched in a range of periods about that predicted by the ephemeris. This range included periods predicted by the more up to date NICER based ephemeris \citep{ygk+20}.

Two X-ray bursts on 2020 May 11 at UTs 02:52:18.000 and 04:22:52.560 with fluences $4.91 \times 10^{-10}$\,erg\,cm$^{-2}$ and $1.35 \times 10^{-8}$\,erg\,cm$^{-2}$, respectively, were detected by \textit{Insight}-HXMT in the MeerKAT observing window (see Table \ref{fig:radioDet} for details). The second of the bursts also fell in the Effelsberg observing window. Another X-ray burst was detected by the \textit{Insight}-HXMT on 2020 May 15 at 00:37:16.000 with a fluence of $8.64 \times 10^{-12}$\,erg\,cm$^{-2}$ which also occurred during the MeerKAT observations. As described above, no pulses were detected at Effelsberg nor at MeerKAT. In Fig.~\ref{fig:fluencelimits} we show a timeline of the single-pulse fluence upper limits at various radio frequencies from this work, together with a selection of measurements reported in the literature. We additionally mark the brightest X-ray bursts detected by \textit{Insight}-HXMT.

Fig.~\ref{fig:meerkat_image} shows the MeerKAT radio image made from the imaging data obtained during the simultaneous beam-formed observations. The radio image shows the SNR~G57.2+00.8 as a complete circular shell with a brightened North-Eastern edge. The enhanced brightness probably indicates that the expanding shock front has encountered dense, molecular interstellar clouds. The high resolution of the image (8.4"$\times$5.8") reveals detailed structure of the bright filaments. The low surface brightness emission towards the South-Western end of the SNR shell shows the two arc-like features also seen by \cite{ksg+18} in the 150 MHz GMRT map. In a region 33$\times$33\,arcsec centred on the known magnetar position we measure a root-mean square flux of 29, 25 and 22\,$\upmu \text{Jy}$\,beam$^{-1}$ for the three imaging epochs (see Table \ref{tab:observations}), respectively. There is clear presence of diffuse emission associated with the SNR located at the position of the magnetar and the corresponding flux values for the pixel located at the source position are 176, 109 and 103 $\upmu \text{Jy}$\,beam$^{-1}$ for the three epochs respectively.

\subsection{Possible radio detection}

A search spanning $\pm 1$ ms around the nominal period and $\pm 200$ pc~cm$^{-3}$ around the nominal DM was carried out on the folded NC data. Observations on 2020 May 30, starting at 00:31:03 UTC and finishing at 02:01:01 UTC, resulted in a marginal detection of a periodic signal with period $P = 3.24760(3)$ s, $\mathrm{DM} = 316(18)$ pc~cm$^{-3}$. The pulse width was of order 100 ms (Fig. \ref{fig:radioDet}). Given the system equivalent flux density of the NC, 640 Jy, this $7-\sigma$ detection implies a flux density of $\sim 3$ mJy. No detections were obtained in any of the other observations, including those performed in the days before and after 2020 May 30, with flux density upper limits for a S/N 6 signal between 2.5 and 5 mJy, depending on the duration of the observation.
However, comparison with the period predicted for this epoch using the ephemeris derived from the NICER observations \citep{ybk+20} indicates a difference in period of approximately 0.2\,ms. As can be seen in the top panel of Figure \ref{fig:radioDet} this offset period lies in the range searched but is clearly not favoured. This therefore suggests that the radio pulse is in fact an unassociated noise fluctuation. During the 35 day NC campaign from 2020 April 30 to 2020 Jun 12, the HXMT reported 6 bursts at UTs of 02:49:27 on 2020-05-02, 01:50:23 and 02:09:32 on 2020-05-16, 01:54:21 on 2020-05-18, 01:24:06 on 2020-05-21 and 00:57:45 on 2020-05-25. No simultaneous radio bursts were detected.

\subsection{Optical MeerLICHT results \label{sec:resultsML}}

No persistent or transient source is detected in the MeerLICHT imaging in any of the exposures and/or filter-bands. Per image 5 $\sigma$ source detection limits are given in Appendix\ \ref{sec:meerlicht} and vary between m$_{AB}$(u) = 18.5 - 19.2, m$_{AB}(g)$=19.3 - 20.4, m$_{AB}$(q)=17.8 - 20.6, m$_{AB}$(r) = 18.9 - 19.8, m$_{AB}$(i) = 18.8 - 19.5 and m$_{AB}$(z) = 17.9 - 18.2. The depth of each image is a combination of image quality (seeing), throughput through the system and background flux (moon in particular). The image quality in the image was at the level of 3$^{\prime\prime}$ which is a combination of site seeing, tracking residuals and a dome seeing due to a sub-optimal airflow through the dome. All non-red-flagged data per filter was co-added for deeper detections. Due to the larger number of observations and the intrinsic sensitivity of the system in the $q$-band this co-add goes substantially deeper than the other bands. No source is detected at the position of SGR\,J1935+2154, averaged over the period May 10 - 19, 2020 down to m$_{AB}(q)$=22.8. Fig.~\ref{fig:meerlicht} shows the full field, and a close-up of the area around the source. The area around SGR\,J1935+2154 shows clear evidence for dust extinction, evidenced by the lower number of star counts over the field. SGR\,J1935+2154 is located at the edge of one of these dust lanes and it is therefore likely that the sight-line to SGR\,J1935+2154 suffers from appreciable extinction. The association of higher star counts with the direct environment of the brighter stars in this region (the brightest of which is HD~184961) shows the correlation of these brighter stars with the dust clouds, i.e. in the direct area of the stars local `holes' are blown in the dust clouds, allowing a clearer line of sight to the dense background star field. Gaia DR2 parallaxes to the brighter stars shown in Fig.~\ref{fig:meerlicht}, having magnitudes $G$=6.3-7.5, imply that the dust clouds are located at only 200-300 pc distance and therefore provide a local absorption screen (see Sect. \ref{sec:discussions}). In the optical there is no sign of the supernova remnant that shows up prominently in the MeerKAT observations (Fig.~\ref{fig:meerlicht}).

\section{Discussion}
\label{sec:discussions}
    Since the discovery of the radio burst, there have been extensive follow-up observations of SGR\,J1935+2154 across the electromagnetic spectrum. The lack of another radio pulse coincident with an X-ray flare puts interesting constraints on the emission mechanism and begs the question of whether we should be able to see such radio bursts in other active Galactic magnetars. In this context, connecting FRBs with extragalactic magnetars is tantalising. Current theories that propose FRB emission from a magnetar can be broadly divided into two categories: 1) far-away models where the FRB is generated by a maser away from the neutron star and 2) close-in models where the FRB is produced in the magnetospehere of the star. The maser emission model runs into difficulties when explaining all the observed radio and X-ray properties of the contemporaneous radio/X-ray burst seen from SGR\,J1935+2154~\citep[see][for more details]{lkz20}.~\cite{ybk+20} have shown that the X-ray burst contemporaneous with the radio burst was spectrally unique compared to all other burst in the activity period and it also supports it having a polar cap origin. Hence, if we assume that FRBs produced by magnetars are created in the magnetosphere close to the polar cap, we can expect them to be significantly beamed \citep{lkz20}. This of course also means that the source must still be exhibiting bursts infrequently in the radio and that more of them might be associated with an X-ray burst than we observe. Observational evidence so far does suggest a connection between the X-ray and the radio emission mechanisms prevalent in neutron stars and any changes in one of them affects the other~\citep{abl+17}. It is believed that while the X-ray bursts and radio pulsations in these sources come from different regions in the magnetosphere, the pair plasma causing the X-ray flares can affect the acceleration of radio emitting particles. This was shown in PSR~J1119--6127, a high B-field radio pulsar where a series of X-ray bursts from the source quenched the radio pulsations with the radio emission returning a few minutes after the last X-ray burst~\citep{abl+17}.
    
\begin{figure*}   
   \centering
   \includegraphics[scale=0.42]{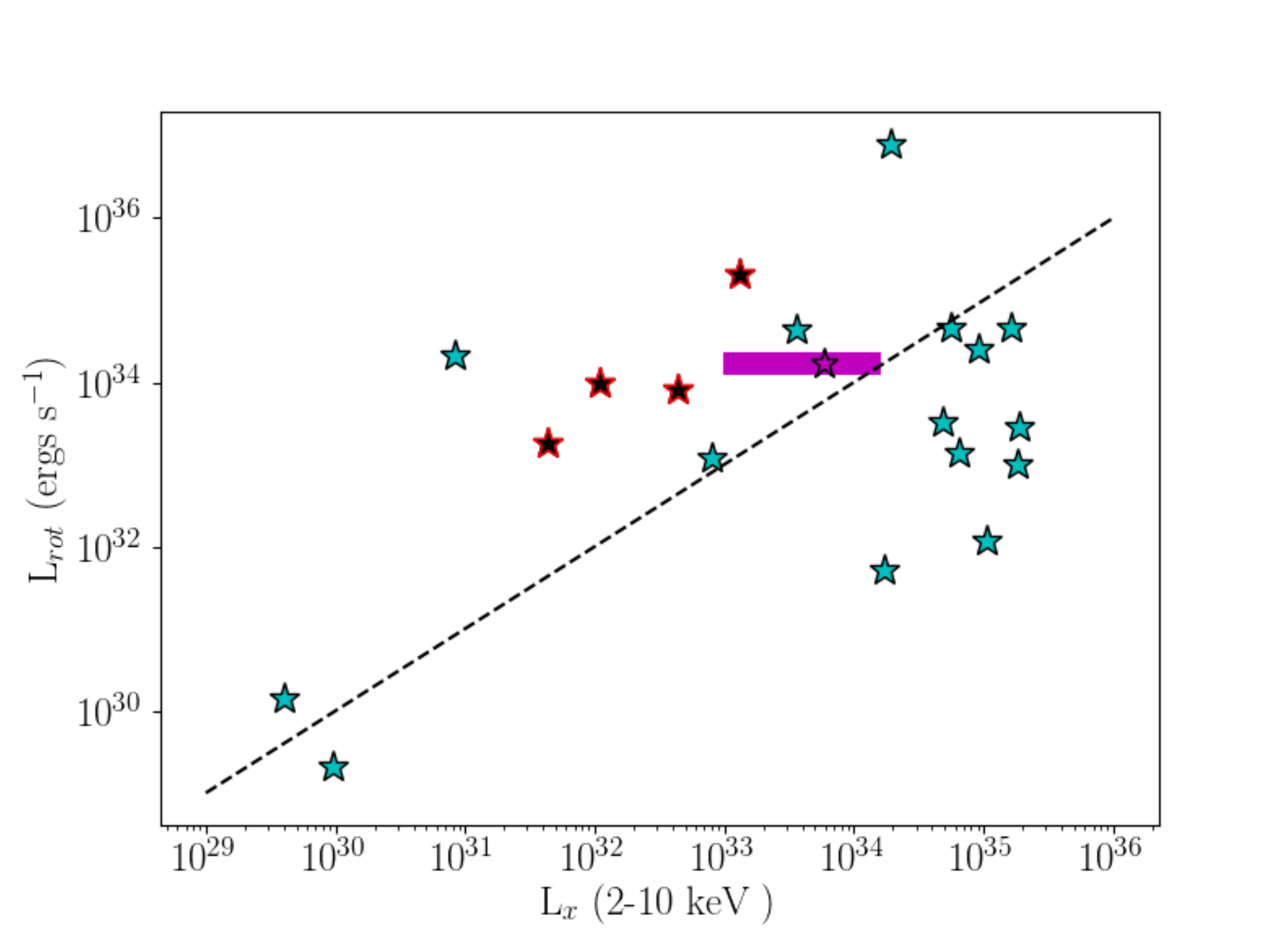}
   \includegraphics[scale=0.42]{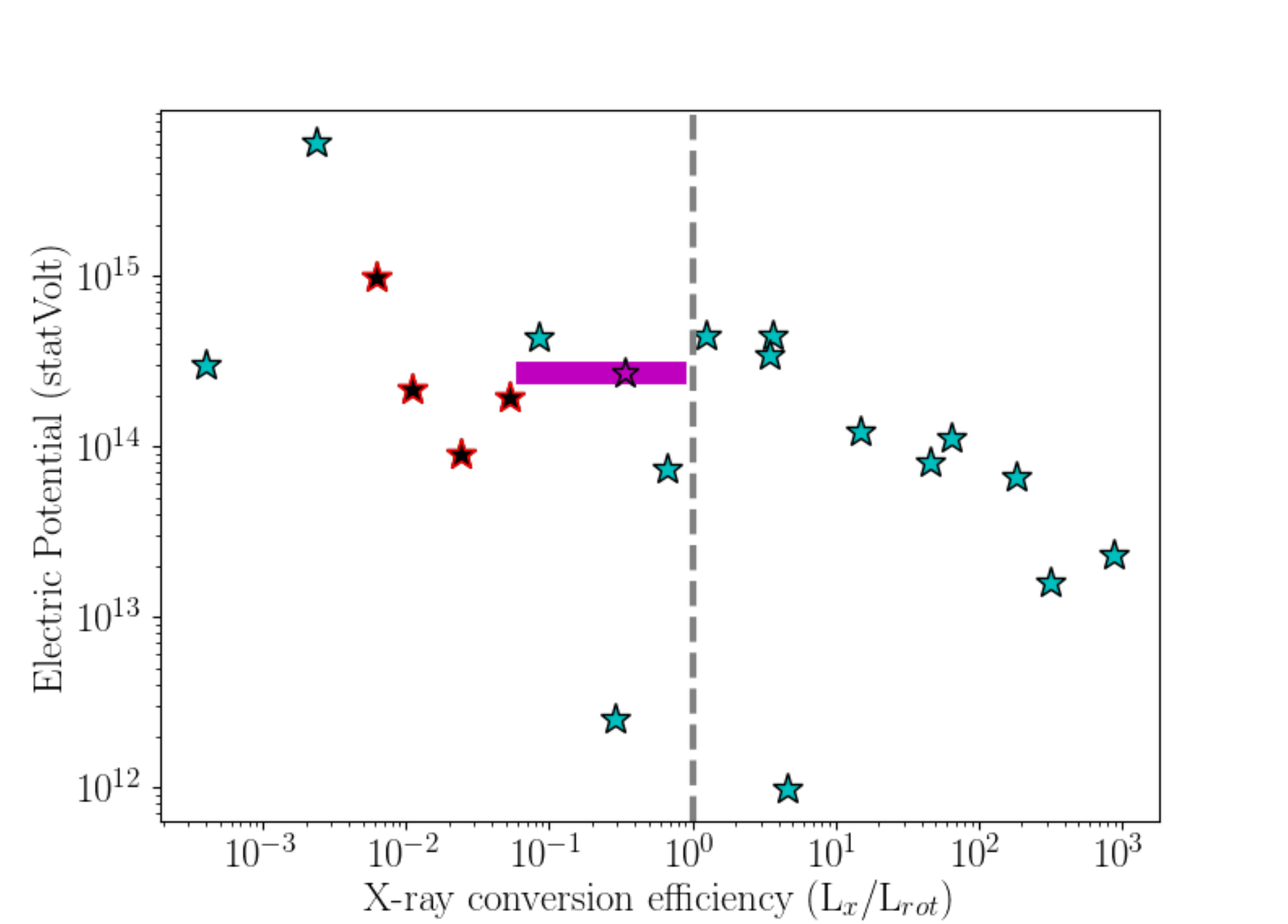}
    \caption{\textbf{Left Panel:} Spin-down Luminosity versus quiescent 2--10~keV X-ray luminosity of magnetars. The black dashed line represents the region where L$_\mathrm{x}$ = L$_\mathrm{rot}$. \textbf{Right Panel:} Electric potential in the magnetosphere as a function of X-ray conversion efficiency. The grey dashed line corresponds to an efficiency of one. For both plots, cyan stars represent the magnetar population taken from the McGill Magnetar Catalog~\citep{magnetarcat}. The black stars with red outlines represent radio loud magnetars while SGR~J1935+2154 is shown as the magenta star. The shaded region represents the range of values L$_\mathrm{x}$ and conversion efficiency based on the distance to the magnetar (see text for more details). The position of the magenta star corresponds to the distance of 6.6~kpc as used in~\citet{bcr+20}.}
    \label{fig:lxlr}
\end{figure*}

    ~\cite{rpt+12} have shown that the radio/X-ray relationship in magnetars and high B-field pulsars can be quantified by three fundamental quantities: 1) the quiescent X-ray luminosity; 2) the spin-down luminosity; and 3) the electric potential. The electric potential that is generated above the polar cap of the neutron star is believed to be responsible for creating the charged plasma responsible for pulsed radio emission~\citep{rs75}. Assuming a dipolar spin-down for a neutron star (which may not be correct for a magnetar), the electric potential,
    \begin{equation}
        V \simeq 4.2\times10^{20}~\sqrt{\frac{P}{\dot{P}^{3}}}\, \rm statvolts,
    \end{equation}
    where $P$ is the period and $\dot{P}$ is the period derivative. In the left panel of Fig.~\ref{fig:lxlr} we show SGR\,J1935+2154 on the so-called fundamental plane of magnetars~\citep{rpt+12} we see an interesting trend. As observed by~\cite{rpt+12}, magnetars that exhibit radio pulsations tend to show a lower X-ray conversion efficiency; the ratio of the quiescent X-ray and spin-down luminosity. The right panel of Fig.~\ref{fig:lxlr} suggests that if the X-ray luminosity of the magnetar is larger than the spin-down luminosity, the electric potential generated in the polar cap reduces. This means that the ability to produce coherent radio pulses might be inversely proportional to the ability of the magnetar to produce X-ray emission~\citep[see][for more details]{rpt+12}. If we assume that the marginal detection from the NC telescope of pulsed radio emission from SGR\,J1935+2154 is genuine, we can associate the ability of the magnetar to produce pulsed radio emission from the magnetic poles to the X-ray efficiency of the source during an active phase. The more efficient a magnetar is in converting the spin-down luminosity and the magnetic energy into X-ray luminosity, the less efficient it is in producing pulsed radio emission. Based on this assumption, SGR\,J1935+2154 lies very close to the transition from efficient to super-efficient regime (L$_{x}$ > L$_{\textrm{rot}}$) and might explain the intermittent nature of the pulsed emission seen from the magnetar so far. However, we note that the bright radio burst seen on April 28$^{\rm th}$~\citep{bkr+20a, scholz2020} and the subsequent detections of single radio pulses apparently emitted within a pulse period over a range of rotational phases~\citep{ksj+20,dg+20} are not consistent with this proposition. One can still argue that the bright single pulses are coming from a different region of the magnetosphere compared to the typical coherent radio emission that we see from radio pulsars. It may also be the case that the radio emission associated with the bright burst is generated by a different physical mechanism or location within the magnetosphere to those seen subsequently with FAST, WSRT and CHIME. This is also consistent with the fact that the X-ray burst that was contemporaneous with the bright radio burst was unlike the typical X-ray bursts seen from the magnetar and constitutes only $\sim$6$\%$ of the X-ray bursts seen from SGR\,J1935+2154~\citep{llx+20,ybk+20,rsf+20}. This emphasizes the need to monitor Galactic magnetars in their active states as they might reveal a peculiar relationship between their radio and X-ray emission properties that will provide more insight into the emission process. 
   
  Radio searches for single pulses and average pulse emission by folding the data based on the ephemeris derived from the X-ray observations \citep[e.g.][]{ier+2016} were carried out using the NC, GMRT \citep{sjm+16} and Arecibo \citep{ykj+18}, on timescales of days to a few weeks after the outbursts in 2014, 2015 and 2016. Neither telescope reported a detection, and the GMRT quotes upper limits of 0.4\,mJy and 0.2\,mJy at 326.5 and 610\,MHz respectively, assuming a 10\% duty cycle and an 8-$\sigma$ detection threshold. The Arecibo observations were at 6.7 and 1.4\,GHz and they report upper limits of 14\,$\mu$Jy and 7\,$\mu$Jy respectively having used a 10-$\sigma$ threshold and a 20\% duty cycle. However, regular monitoring is supported by the most recent detection of just 3 pulses with CHIME \citep{dg+20} and the subsequent detection of single pulses and average pulse emission with FAST \citep{zb+20} which also appears to have been transient in nature. 
  It is also unclear how long it takes before the radio emission turns on in these systems. Sometimes it is relatively rapid \citep[e.g.][]{lld+19} and other times it can be quite some time after the X-ray outbursts~\citep[see][for more details]{kb17} and can apparently also vary for a given source too. Usually when a magnetar turns on in the radio, it emits radio pulses during most rotations and stays on, with some significant variability in flux density and pulse shape, over a timescale of months to years. In the case of SGR\,J1935+2154, the~\cite{ksj+20} campaign lasted for 2 months with a total observing time of 522 hours while our campaign lasted for 1.5 months for a total integration time of $\sim$110 hours. Between these two overlapping campaigns, only two radio pulses were seen by~\cite{ksj+20}. These results along with the the very recent detection of bright radio single pulses and pulsed radio emission from the magnetar~\citep{dg+20,zb+20} suggest that the change in the magnetosphere configuration in SGR\,J1935+2154 to enable pulsed radio emission might be more gradual or sporadic than other magnetars like XTE~J1810--197 and Swift~J1818.0--1607 where bright radio pulses were seen in the immediate aftermath of the X-ray bursting behaviour. A possible change in the magnetosphere is also supported by X-ray pulsar variations \citep{ygk+20}. This diversity of emission characteristics emphasizes the need to regularly monitor magnetars in their active states. 
  
   The MeerKAT images show no evidence of a point source at the location of SGR\,J1935+2154 during any of the four observing epochs (including those taken during the observations on the 29th which also resulted in a non-detection). As mentioned above we quote flux limits only for the first 3 observing sessions as there was no calibrator scan available for the initial observation which was taken during an ongoing pulsar timing observation. The flux limits are dominated by the presence of the diffuse emission from the SNR. However, they are comparable to those obtained in observations made at similar frequencies with the uGMRT (see Table~\ref{fig:radioDet}) although somewhat worse as the uGMRT resolves out more of the emission. This is why the limit from the European VLBI Network telescopes is also a factor of a few fainter. There was no detection of any persistent or transient optical emission seen with MeerLICHT. The limits from MeerLICHT are in line with the faint infrared counterpart (F140W(AB)=25.3) reported by \cite{lkf18}, given the extinction along the line of sight and any reasonable slope on the source's spectrum between the optical and the nIR.
   
\begin{figure}
  \centering
  \includegraphics[width=9cm]{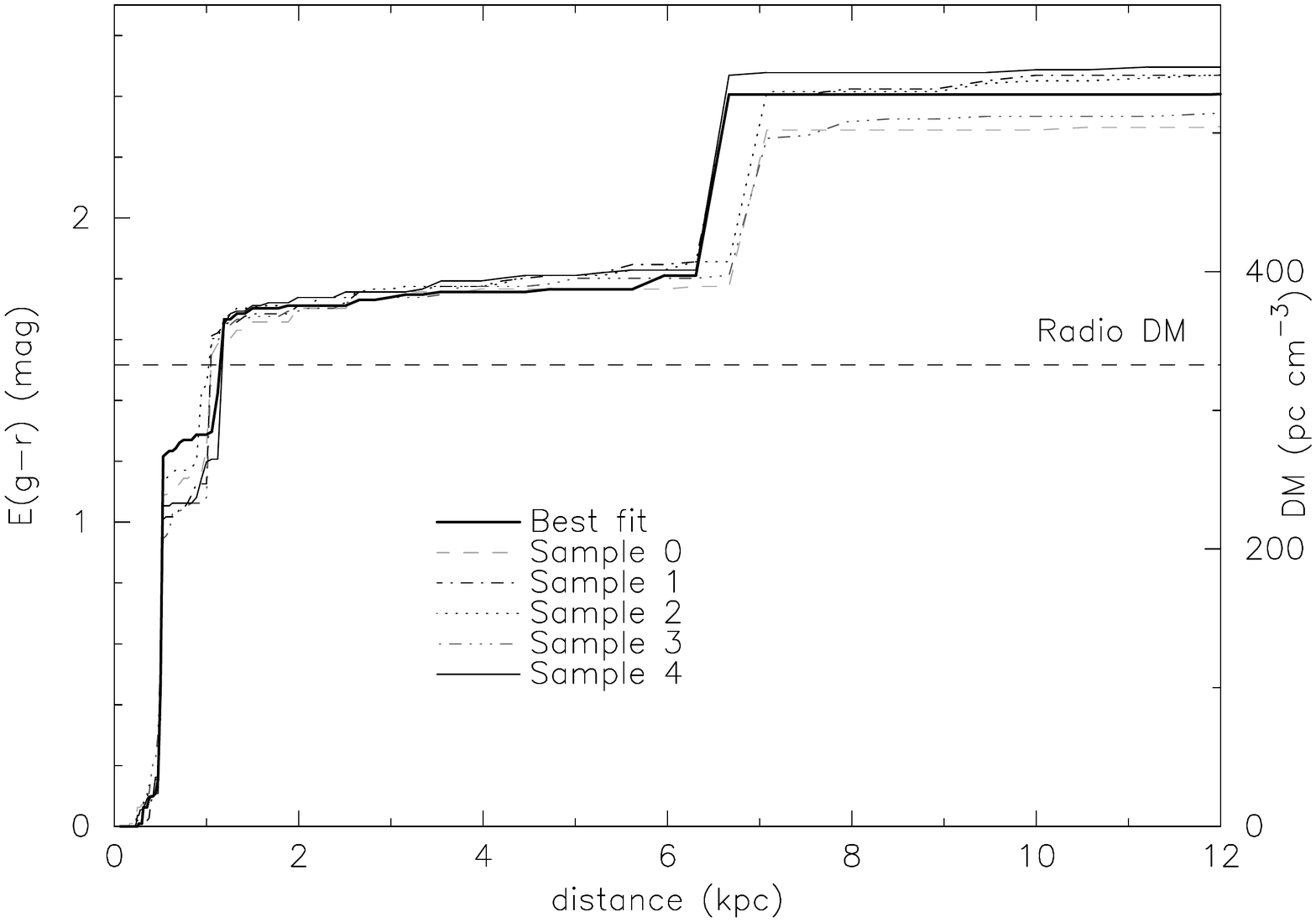}
  \caption{Colour-excess $E(g-r)$ (left axis) and dispersion measure (DM, right axis) as a function of distance, derived from optical data, for the position of SGR\,J1935+2154, as extracted from the data of \citet{gsz19}. Thin lines show the various sample data sets and the thick lines the best fit from \citet{gsz19}. To calculate the dispersion measure Eq.~\ref{eq:DM-E} has been used.}
  \label{fig:extinction}
\end{figure}
   
   The distance and, perhaps also, the association of SGR\,J1935+2154 with the SNR G57.2+00.8 is still debatable. This was most recently noted in the paper by \cite{zdz+20} who pointed out that the distance estimates can be divided into two classes, those that are based on the SNR itself and those that are based on the magnetar. Radio based studies looking either at the remnant itself or {\ion HI} and $N_{\rm H}$ and dispersion measure models\footnote{before the magnetar was detected as a radio source} give distances in the range of 4.5 to 14\,kpc \citep{puv+13,sjm+16,ksg+18,rlt+18} while a recent study of CO and molecular clouds in the vicinity of the remnant gives a distance of 6.6$\pm$0.7 kpc \citep{zzc+20}. Based on a blackbody spectral fit to the emission of an X-ray flare in a previous outburst \cite{kis+16} derive a distance upper limit of 10\,kpc while \cite{msf+20} use their detection of an expanding dust scattering halo during the most recent outburst to constrain the distance to a range 2.2 $-$ 7.1\,kpc. In their paper \cite{zdz+20} determine a distance of 9.0$\pm$2.5\,kpc based mainly on the YMW16 \citep{ymw16} dispersion measure based distance model.

    With these wide range of distances we note, that although not very accurate, it is possible to put constraints on the distance to SGR\,J1935+2154 from its detected DM and a relation that links DM to distance within the Milky Way Galaxy. This relation ties together a number of steps, starting with a 3D extinction map within the Galactic Plane. From a colour excess -- distance relation ($E(g-r)$ vs. $d$), an absorption ($A_{\rm V}$) and colour excess relation, an absorption - hydrogen column density ($N_{\rm H}$) relation and a hydrogen column density -- dispersion measure relation it is possible to link dispersion measure to distance. The first step is to use 3D dust mapping as in \cite{gsz19}, \cite{sdu09}, \cite{sale14}. Here we use the results from \cite{gsz19} to link the detected colour excess ($E(g-r)$) to distance. Using Green et al.'s \href{http://argonaut.skymaps.info/query?lon=57.24664636%20%20%20&lat=%20%200.81900950%20&coordsys=gal&mapname=bayestar2019}{
online webtool}, the derived relation at the position of SGR\,J1935+2154 is shown in Fig.\ \ref{fig:extinction}. The colour-excess in this direction is due to a number of discrete absorption screens, located at a distance ($d$) of roughly $d$=500 pc, $d$=1.5 kpc and $d$=6.5 kpc. The bright stars and dust clouds shown in Fig. \ref{fig:meerlicht} are clearly part of the first of these absorption screens. The colour-excess is converted to absorption magnitudes in the Johnson V-band, $A_{\rm V}$ using equation 30 from \cite{gsz19} : 
   \begin{equation}A_{\rm V} = (E(g-r)-C_1)/C_2, \label{eq:Av}\end{equation} 
   where $C_1$ = 0.03 and $C_2$ = 0.269. The conversion from absorption $A_{\rm V}$ to hydrogen column density ($N_{\rm H}$) is derived from X-ray studies on scattering halos or X-ray spectral fitting to supernova remnants and takes the form 
   \begin{equation}N_{\rm H} = C_3 A_{\rm V}, \label{eq:NH}\end{equation} 
   where the exact value of $C_3$ differs between studies, e.g. \cite{g75}; \cite{ps95}; \cite{go09}; \cite{tsx13}. 
   The relation between hydrogen column density and dispersion measure is given by \cite{hnk13} as
   \begin{equation}{\rm DM} =N_{\rm H}/C_4, \label{eq:DM}\end{equation} 
   where $C_4$ = \gpm{0.030}{0.013}{0.009}, when $N_{\rm H}$ is expressed in units of 10$^{21}$ cm$^{-2}$.  
   Combining Eqs.\ \ref{eq:Av}, \ref{eq:NH} and \ref{eq:DM} leads to the relation: 
   \begin{equation}
   {\rm DM} = \frac{C_3 (E(g-r) - C_1)}{C_2 C_4}.
   \label{eq:DM-E}
   \end{equation}
   The final step of relating dispersion measure to distance results from the empirical relation shown in Fig.\ \ref{fig:extinction}. Using the \cite{ps95} value of $C_3 = 1.79\pm{0.03}$ and evaluating the other constants at their nominal values shows that the colour-excess plateau in Fig.\ \ref{fig:extinction} at 1.65 $< E(g-r)<$ 1.83, corresponding to a distance between 1.2 kpc $<d<$ 6 kpc, leads to a dispersion measure range 360 pc\,cm$^{-3}<$\,DM\,$<$ 400 pc\,cm$^{-3}$. The further plateau in extinction at levels of $E(g-r) \geq 2.4$ leads to DM $\geq$ 500 pc cm$^{-3}$ (see Fig.\ \ref{fig:extinction}). These results are consistent with the results presented by \cite{sale14} based on the IPHAS survey \citep{drew05} covering the same coordinates, where the relation is given between distance and monochromatic absorption A$_0$ evaluated at a wavelength of 5459 \AA, corresponding to the central wavelength of the Johnson V-band. Given the uncertainties and the fact that $N_{\rm H}$ values determined for magnetars close to the Galactic plane tend to be higher due to X-ray absorption by molecular species rather than just atomic species \citep{hnk13}, the value of $N_{\rm H} = 1.1\times10^{22}$\,cm$^{-2}$ derived from the reddening is consistent with the values determined from spectral fitting by \cite{ier+2016} of 2 $\pm~0.4\times10^{22}$\,cm$^{-2}$. 
   
   Translating the measured DM$ \approx 333$\,pc\,cm$^{-3}$ directly to distance using this method places the source inside the second absorption screen at a close distance of $d\sim$1.2 kpc. Given the large uncertainties associated with the constants in Eq.\ \ref{eq:DM-E} an association on the colour excess plateau between 1.2 kpc $< d <$ 6.5 kpc is perhaps more likely. Any dispersion measure contribution originating locally to the source would lower the amount of dispersion measure attributable to the interstellar medium and would potentially favour the closer distance. However we note that \cite{zdz+20} suggest that the contribution from the remnant should be small. The upper limit of the reddening derived distance range, about 6.5\,kpc is most consistent with other distance estimates based on the magnetar itself, in particular that from the dust scattering halo (2.2 $-$ 7.1\,kpc; \cite{msf+20} and also with the distance of the SNR determined from the molecular clouds (6.6$\pm$0.7 kpc; \cite{zzc+20}). However 6.5\,kpc is at the lower limit of the other distances derived based on methods that use properties or the line of sight to the SNR. The DM of the detected radio pulses does provide a constraint on the distance using models of the dispersion measure but the uncertainty in the DM-derived distance is always fairly large \citep[see][for a discussion on the distances]{zdz+20} especially given that the optical observations suggest that there may significant material at just a few hundred parsecs. 
   
   The significant step up in DM suggested from the reddening as shown in Fig. \ref{fig:extinction} combined with the other measurements means that we favour a distance of around 6.5\,kpc for the magnetar. This distance does have some tension with the preferred distances to the SNR but they are not completely inconsistent and therefore do not necessarily imply that the association between SGR~J1935+2154 and SNR G57.2+00.8 is in doubt. However they do argue for a future more detailed study of the association. A distance of 6.5\,kpc is nearer than the mean value of around 9 -- 10\,kpc that has been used in many papers \citep[e.g.][]{puv+13,kis+16,rlt+18,zdz+20,kis+16,msf+20,zzc+20} for determining the luminosity of the initial bright burst detected on 2020 April 28. In their estimates of the energy/luminosity of the April 28 burst CHIME \cite{abb+20} and STARE2 used distances of 10 and 9.5\,kpc respectively. If we use our nearer distance of 6.5\,kpc this would decrease the energy of the burst by about a factor of two and therefore suggest that it is closer to two orders of magnitude fainter than the faintest extragalactic FRBs seen so far. It does however still make it at least a couple of orders of magnitude brighter than anything seen in the Galaxy from a neutron star.

\section{Conclusions}
\label{sec:conclusion}
We have presented results from a radio and optical monitoring campaign of SGR~J1935+2154 during its active state when it emitted a bright radio burst contemporaneous with an X-ray flare. We observed the source during a number of campaigns which included several simultaneous multi-frequency campaigns coincident with X-ray observations as well as longer term monitoring campaigns with a range of telescopes at different frequencies. We failed to detect any single pulses of radio emission despite having observations near in time to the detections made by the EVN project \citep{ksj+20}. A tentative detection of average pulse emission with the NC which was quite close in time to non-detections of average pulse emission and single pulses with other telescopes was ruled out due to the difference in period compared to that derived from X-ray observations. The range of pulse energies at which single pulse emission has been detected, and for apparently brief times, suggests that the source is exceptionally variable, even given the highly variable nature of radio emission seen from the other Galactic magnetars.  The location of SGR J1935+2154 in the fundamental radio--X-ray plane for magnetars suggests that the object is on the threshold of how efficiently it can produce radio emission and this might be related to the variety of emission that we see in the radio. Some of the radio observations were coincident with X-ray bursts and so it confirms the model that not all X-ray bursts have associated radio emission, potentially due to geometry, but also the fact that the X-ray burst associated with the April 28 event was significantly different.

We find that there is no persistent or transient radio emission at the X-ray position of the magnetar in the MeerKAT radio images above a best limit of 103~$\mu$Jy beam$^{-1}$ indicating that there is no magnetar wind nebula formed which is radio emitting at our detection threshold. There is also no optical emission detected during the period of observation. The optical observations of the field combined with the DM of the magnetar allowed us to obtain a distance estimate for the magnetar which support a closer distance. This would suggest that the FRB-like burst might be a factor of 2 or more less luminous than previously thought and thus about two orders of magnitude fainter than the least luminous of the known extragalactic FRB pulses. It also suggests that a more detailed study of the association with SNR G57.2+00.8 is needed. This is important as the magnetar models for extra-galactic FRBs is based on them being associated with young magnetars and so one might expect that there would be an associated SNR. The only occasional detections of radio pulses highlighted by our, and other, observing campaigns demonstrates the need of regular multi-wavelength follow-up of active magnetars as this may lead us to a better understanding the origins of FRBs.

\section*{Acknowledgements}

The authors would like to thank SARAO for the approval of the MeerKAT DDT request and the CAM/CBF/SP and operator teams for their time and effort invested in the observations. The MeerKAT telescope is operated by the South African Radio Astronomy Observatory (SARAO), which is a facility of the National Research Foundation, an agency of the Department of Science and Innovation. This project has received funding from the European Research Council (ERC) under the European Union’s Horizon 2020 research and innovation programme (grant agreement No. 694745). We acknowledge support of the Gravitational Wave Data Centre funded by the Department of Education via Astronomy Australia Ltd and the Australian Research Council Centre of Excellence for Gravitational Wave Discovery (OzGrav), through project number CE170100004. Part of this work has been funded using resources from the research grant {\it iPeska} (P.I. Andrea Possenti) funded under the INAF national call Prin-SKA/CTA approved with the Presidential Decree 70/2016. AP and MBu acknowledge the support from the Ministero degli Affari Esteri e della Cooperazione Internazionale - Direzione Generale per la Promozione del Sistema Paese - Progetto di Grande Rilevanza ZA18GR02.\\
The MeerLICHT consortium is a partnership between Radboud University, the University of Cape Town, the Netherlands Organisation for Scientific Research (NWO), the South African Astronomical Observatory (SAAO), the University of Oxford, the University of Manchester and the University of Amsterdam, in association with and, partly supported by, the South African Radio Astronomy Observatory (SARAO), the European Research Council and the Netherlands Research School for Astronomy (NOVA). We acknowledge the use of the Inter-University Institute for Data Intensive Astronomy (IDIA) data intensive research cloud for data processing. IDIA is a South African university partnership involving the University of Cape Town, the University of Pretoria and the University of the Western Cape. This paper is based (in part) on data obtained with the International LOFAR Telescope (ILT) under project code DDT13\_006. LOFAR is the Low Frequency Array designed and constructed by ASTRON. It has observing, data processing, and data storage facilities in several countries, that are owned by various parties (each with their own funding sources), and that are collectively operated by the ILT foundation under a joint scientific policy. The ILT resources have benefitted from the following recent major funding sources: CNRS-INSU, Observatoire de Paris and Universit\'e d'Orl\'eans, France; BMBF, MIWF-NRW, MPG, Germany; Science Foundation Ireland (SFI), Department of Business, Enterprise and Innovation (DBEI), Ireland; NWO, The Netherlands; The Science and Technology Facilities Council, UK. The Arecibo Observatory is operated by the University of Central Florida, Ana G. Mendez-Universidad Metropolitana, and Yang Enterprises under a cooperative agreement with the National Science Foundation (NSF; AST-1744119). GD, KL, MK and RK are supported by the European Research Council  Synergy Grant BlackHoleCam under contract no.\ 610058. This publication is based on observations with the 100-m telescope of the Max-Planck-Institut für Radioastronomie at Effelsberg. We would like to thank the referee for the careful reading of the manuscript.  

%%%%%%%%%%%%%%%%%%%%%%%%%%%%%%%%%%%%%%%%%%%%%%%%%%



\bibliographystyle{mnras}
\bibliography{sgr1935} % if your bibtex file is called example.bib

%\begin{thebibliography}{99}
%\end{thebibliography}

%%%%%%%%%%%%%%%%%%%%%%%%%%%%%%%%%%%%%%%%%%%%%%%%%%
% \onecolumn
\newpage
\vspace{0.2cm}
% \centering
% \scriptsize
\noindent  \textit{% List of institutions
$^{1}$Centre for Astrophysics and Supercomputing, Swinburne University of Technology,
Mail 74 PO Box 218, Hawthorn, Vic, 3122\\
$^{2}$ARC Centre of Excellence for Gravitational Wave Discovery (OzGrav)\\
$^{3}$ASTRON, Netherlands Institute for Radio Astronomy, Oude
Hoogeveensedijk 4, 7991 PD, Dwingeloo, The Netherlands\\
$^{4}$INAF-Istituto di Radio Astronomia, via Gobetti 101, 40129 Bologna, Italy\\
$^{5}$Department of Physics and Electronics, Rhodes University, PO Box 94, Grahamstown, 6140, South Africa\\
$^{6}$South African Radio Astronomy Observatory, Black River Park, 2 Fir Street, Observatory, Cape Town, 7925, South Africa\\
$^7$Jodrell Bank Centre for Astrophysics, Department of Physics and Astronomy, University of Manchester, Manchester M13 9PL, UK\\
$^{8}$Anton Pannekoek Institute for Astronomy, University of
Amsterdam, Science Park 904, 1098 XH Amsterdam, The Netherlands\\
$^9$Max-Planck-Institut f\"ur Radioastronomie, Auf dem H\"ugel 69, D-53121 Bonn, Germany\\
$^{10}$LESIA, Observatoire de Paris, Universit\'e PSL, CNRS, Sorbonne Universit\'e, Universit\'e de Paris, 5 place Jules Janssen, 92195 Meudon, France\\
$^{11}$Department of Astrophysics/IMAPP, Radboud University, P.O. 9010,
6500 GL, Nijmegen, The Netherlands\\ 
$^{12}$Inter-University Institute for Data Intensive Astronomy \&
Department of Astronomy, University of Cape Town, Private Bag X3,
Rondebosch 7701, South Africa\\ 
$^{13}$South African Astronomical Observatory, P.O. Box 9, 7935
Observatory, South Africa\\
$^{14}$Astrophysics, Department of Physics, University of Oxford, Keble Road, Oxford, OX1 3RH, UK\\ 
$^{15}$INAF-Osservatorio Astronomico di Cagliari, via della Scienza 5, I-09047 Selargius (Cagliari), Italy\\
$^{16}$INAF OAS Bologna, Via P. Gobetti 101, 40129 Bologna, Italy \\
$^{17}$Astro Space Centre, Lebedev Physical Institute, Russian Academy of Sciences, Profsoyuznaya Str. 84/32, Moscow 117997, Russia\\
$^{18}$Leiden Observatory, Leiden
University, P.O. Box 9513, NL-2300 RA Leiden, The Netherlands\\
$^{19}$CSIRO Astronomy and Space Science, Australia Telescope National Facility, PO Box 76, Epping NSW 1710, Australia\\
$^{20}$Institute of Space Sciences and Astronomy (ISSA), University of Malta, Msida MSD 2080, Malta\\
$^{21}$IAU-Office For Astronomy for Development, P.O. Box 9, 7935
Observatory, South Africa\\ 
$^{22}${Center for Interdisciplinary Exploration and Research in
Astrophysics (CIERA), Northwestern University, 1800 Sherman Ave,
Evanston, IL 60201, USA}\\ 
$^{23}$Arecibo Observatory, Route 625 Bo. Esperanza, Arecibo, PR 00612, USA \\
$^{24}$Department of Physics, McGill University, 3600 University
Street, Montr\'{e}al, QC H3A 2T8, Canada\\
 $^{25}$McGill Space Institute, McGill University, 3550 University
Street, Montr\'{e}al, QC H3A 2A7, Canada\\
$^{26}$Università degli Studi di Cagliari, Dip di Fisica, S.P. Monserrato-Sestu Km 0,700, I-09042 Monserrato (CA) , Italy
$^{27}$Dipartimento di Fisica e Astronomia, Universit\'{a} di Bologna, Via Gobetti 93/2, 40129 Bologna, Italy\\
$^{28}$INAF/IAPS, via del Fosso del Cavaliere 100, I-00133 Roma (RM), Italy\\
$^{29}$Università degli Studi di Roma "Tor Vergata", via della Ricerca Scientifica 1, I-00133 Roma (RM), Italy\\
}

%%%%%%%%%%%%%%%%% APPENDICES %%%%%%%%%%%%%%%%%%%%%

\appendix

\section{Radio Monitoring Campaign}

\begin{table*}
	\centering
	\caption{Radio monitoring campaign of SGR\,J1935+2154 using the Medicina Northern Cross Telescope and the MK2 telescope at Jodrell Bank Observatory.}
	\label{tab:ncMK2obs}
	\begin{tabular}{lcccccc} % four columns, alignment for each
		\hline
		Telescope & T$_{\mathrm{start}}$ & T$_{\mathrm{end}}$& F$_{\mathrm{centre}}$ & BW & N$_{\mathrm{chan}}$ & t$_{\mathrm{samp}}$\\
	             &	(UTC) & (UTC) & (MHz) & (MHz) & & ($\mathrm{\mu s}$)\\
	    \hline
Northern Cross & 2020-04-30 03:00:25 & 2020-04-30 03:31:41 & 408 & 16 & 1311 & 138.24 \\
Northern Cross & 2020-05-02 02:23:20 & 2020-05-02 03:53:01 & 408 & 16 & 1311 & 138.24 \\
Northern Cross & 2020-05-03 02:20:47 & 2020-05-03 03:50:27 & 408 & 16 & 1311 & 138.24 \\
Northern Cross & 2020-05-04 02:15:24 & 2020-05-04 03:45:01 & 408 & 16 & 1311 & 138.24 \\
Northern Cross & 2020-05-05 02:33:21 & 2020-05-05 03:45:13 & 408 & 16 & 1311 & 138.24 \\
Northern Cross & 2020-05-06 02:07:24 & 2020-05-06 03:37:01 & 408 & 16 & 1311 & 138.24 \\
Northern Cross & 2020-05-12 01:43:20 & 2020-05-12 02:10:00 & 408 & 16 & 1311 & 138.24 \\
Northern Cross & 2020-05-14 01:36:03 & 2020-05-14 02:10:56 & 408 & 16 & 1311 & 138.24 \\
Northern Cross & 2020-05-15 01:39:08 & 2020-05-15 03:09:06 & 408 & 16 & 1311 & 138.24 \\
Northern Cross & 2020-05-16 01:27:03 & 2020-05-16 02:09:42 & 408 & 16 & 1311 & 138.24 \\
Northern Cross & 2020-05-17 01:23:03 & 2020-05-17 02:06:15 & 408 & 16 & 1311 & 138.24 \\
Northern Cross & 2020-05-18 01:19:03 & 2020-05-18 02:49:01 & 408 & 16 & 1311 & 138.24 \\
Northern Cross & 2020-05-19 01:15:03 & 2020-05-19 02:45:01 & 408 & 16 & 1311 & 138.24 \\
Northern Cross & 2020-05-20 01:10:51 & 2020-05-20 02:40:49 & 408 & 16 & 1311 & 138.24 \\
Northern Cross & 2020-05-21 01:06:50 & 2020-05-21 02:36:48 & 408 & 16 & 1311 & 138.24 \\
Northern Cross & 2020-05-22 01:03:03 & 2020-05-22 02:33:01 & 408 & 16 & 1311 & 138.24 \\
Northern Cross & 2020-05-23 00:59:02 & 2020-05-23 02:29:00 & 408 & 16 & 1311 & 138.24 \\
Northern Cross & 2020-05-24 00:55:02 & 2020-05-24 02:25:00 & 408 & 16 & 1311 & 138.24 \\
Northern Cross & 2020-05-25 00:51:02 & 2020-05-25 02:21:01 & 408 & 16 & 1311 & 138.24 \\
Northern Cross & 2020-05-26 00:47:02 & 2020-05-26 02:17:00 & 408 & 16 & 1311 & 138.24 \\
Northern Cross & 2020-05-27 00:43:02 & 2020-05-27 02:13:00 & 408 & 16 & 1311 & 138.24 \\
Northern Cross & 2020-05-28 00:39:01 & 2020-05-28 02:09:00 & 408 & 16 & 1311 & 138.24 \\
Northern Cross & 2020-05-29 00:35:01 & 2020-05-29 02:05:00 & 408 & 16 & 1311 & 138.24 \\
Northern Cross & 2020-05-30 00:31:03 & 2020-05-30 02:01:01 & 408 & 16 & 1311 & 138.24 \\
Northern Cross & 2020-05-31 00:27:02 & 2020-05-31 01:57:01 & 408 & 16 & 1311 & 138.24 \\
Northern Cross & 2020-06-01 00:23:03 & 2020-06-01 01:53:01 & 408 & 16 & 1311 & 138.24 \\
Northern Cross & 2020-06-02 00:19:02 & 2020-06-02 01:49:00 & 408 & 16 & 1311 & 138.24 \\
Northern Cross & 2020-06-03 00:15:02 & 2020-06-03 01:45:00 & 408 & 16 & 1311 & 138.24 \\
Northern Cross & 2020-06-04 00:11:03 & 2020-06-04 01:41:00 & 408 & 16 & 1311 & 138.24 \\
Northern Cross & 2020-06-05 00:07:02 & 2020-06-05 01:37:00 & 408 & 16 & 1311 & 138.24 \\
Northern Cross & 2020-06-07 23:56:02 & 2020-06-08 01:26:00 & 408 & 16 & 1311 & 138.24 \\
Northern Cross & 2020-06-08 23:52:02 & 2020-06-09 01:22:00 & 408 & 16 & 1311 & 138.24 \\
Northern Cross & 2020-06-09 23:48:02 & 2020-06-10 01:18:00 & 408 & 16 & 1311 & 138.24 \\
Northern Cross & 2020-06-11 23:42:03 & 2020-06-12 01:12:01 & 408 & 16 & 1311 & 138.24 \\
Northern Cross & 2020-06-12 23:37:02 & 2020-06-13 01:07:01 & 408 & 16 & 1311 & 138.24 \\
MK2 & 2020-05-31 07:03:16 & 2020-05-31 07:58:16 & 1532 & 336 & 1538 & 256 \\
MK2 & 2020-06-01 06:59:21 & 2020-06-01 07:59:16 & 1532 & 336 & 1538 & 256 \\
MK2 & 2020-06-01 08:11:13 & 2020-06-01 08:58:40 & 1532 & 336 & 1538 & 256 \\
MK2 & 2020-06-04 07:40:32 & 2020-06-04 08:35:00 & 1532 & 336 & 1538 & 256 \\
MK2 & 2020-06-04 07:03:16 & 2020-06-04 07:58:16 & 1532 & 336 & 1538 & 256 \\
MK2 & 2020-06-05 07:45:32 & 2020-06-05 08:32:30 & 1532 & 336 & 1538 & 256 \\
MK2 & 2020-06-06 06:39:44 & 2020-06-06 07:39:44 & 1532 & 336 & 1538 & 256 \\
MK2 & 2020-06-06 07:41:36 & 2020-06-06 08:27:36 & 1532 & 336 & 1538 & 256 \\
MK2 & 2020-06-08 06:31:55 & 2020-06-08 07:31:55 & 1532 & 336 & 1538 & 256 \\
MK2 & 2020-06-08 07:33:48 & 2020-06-08 08:07:48 & 1532 & 336 & 1538 & 256 \\
MK2 & 2020-06-09 06:27:54 & 2020-06-09 07:27:54 & 1532 & 336 & 1538 & 256 \\
MK2 & 2020-06-09 07:29:47 & 2020-06-09 08:15:47 & 1532 & 336 & 1538 & 256 \\
MK2 & 2020-06-10 06:24:05 & 2020-06-10 07:24:05 & 1532 & 336 & 1538 & 256 \\
MK2 & 2020-06-10 07:26:00 & 2020-06-10 08:12:00 & 1532 & 336 & 1538 & 256 \\
MK2 & 2020-06-11 06:20:21 & 2020-06-11 07:20:21 & 1532 & 336 & 1538 & 256 \\
		\hline
	\end{tabular}
\end{table*}

\section{MeerLICHT observations} 
\label{sec:meerlicht}

\begin{table*}
	\centering
	\caption{Overview of all MeerLICHT observations on the fieldID 11744 containing the position of SGR~1935+2154, ordered by filter and date. All exposures were 60~s in duration. Limiting magnitudes/fluxes for the detection of persistent sources are at a signal-to-noise ratio of 5$\sigma$; limiting magnitudes/fluxes for the detection of transient sources are at a transient peak significance level (Scorr value) of 6. Red flagged images do not get processed for transient detections.}
	\label{tab:meerlicht}
	\begin{tabular}{rrrrrrrrrr} % four columns, alignment for each
		\hline
        Image ID & Image Name & Filter & Date & UT-mid & Flag & Lim. Mag & Lim. Flux & Lim Mag & Lim. Flux \\
		         & & & & & & Persistent & Persistent & Transient & Transient\\ 
		         & & & & & & (AB) & ($\mu$Jy) & (AB) & ($\mu$Jy)\\ \hline

46166&	ML1\_20200510\_022440\_red&	$u$&	2020-05-10&	02:24:41.41&	red   &18.53&	141.19&	-    & -		 \\
46172&	ML1\_20200510\_023712\_red&	$u$&	2020-05-10&	02:37:13.13&	red   &18.58&	133.90&	-    &-		 \\
46178&	ML1\_20200510\_024940\_red&	$u$&	2020-05-10&	02:49:41.41&	red   &18.71&	119.05&	-    &-		 \\
46948&	ML1\_20200515\_024948\_red&	$u$&	2020-05-15&	02:49:48.48&	red   &19.14&	80.54 &	-    &-	         \\
47221&	ML1\_20200517\_024948\_red&	$u$&	2020-05-17&	02:49:48.48&	red   &19.09&	84.00 &	-    &-		 \\
47384&	ML1\_20200518\_025240\_red&	$u$&	2020-05-18&	02:52:41.41&	red   &19.27&	70.99 &	-    &-		 \\
47469&	ML1\_20200519\_025027\_red&	$u$&	2020-05-19&	02:50:27.27&	red   &18.70&	119.70&	-    &-		 \\
48101&	ML1\_20200521\_030621\_red&	$u$&	2020-05-21&	03:06:21.21&	red   &18.73&	117.02&	-    &-		 \\
     &                         &         &                &                &          &     &         &      &           \\ 	           			        	  		   		    	             	       	      		 
46167&	ML1\_20200510\_022645\_red&	$g$&	2020-05-10&	02:26:46.46&	yellow&19.35&	66.17 &	-    &-	         \\
46173&	ML1\_20200510\_023918\_red&	$g$&	2020-05-10&	02:39:19.19&	yellow&19.35&	66.26 &	18.78&	111.79	 \\
46179&	ML1\_20200510\_025146\_red&	$g$&	2020-05-10&	02:51:47.47&	yellow&19.29&	69.81 &	18.75&	114.77	 \\
46949&	ML1\_20200515\_025154\_red&	$g$&	2020-05-15&	02:51:54.54&	yellow&19.98&	36.82 &	18.98&	92.55 	 \\
47222&	ML1\_20200517\_025153\_red&	$g$&	2020-05-17&	02:51:53.53&	yellow&20.20&	30.25 &	19.04&	87.79  	 \\
47385&	ML1\_20200518\_025444\_red&	$g$&	2020-05-18&	02:54:45.45&	yellow&20.37&	25.93 &	19.07&	85.37 	 \\
47470&	ML1\_20200519\_025232\_red&	$g$&	2020-05-19&	02:52:32.32&	red   &19.41&	62.80 &	-    &-	       	 \\
48102&	ML1\_20200521\_030826\_red&	$g$&	2020-05-21&	03:08:26.26&	yellow&19.90&	39.92 &	18.93&	97.47 	 \\
     &                         &         &                &                &          &     &         &      &           \\ 			        	  		   		    	             	       	      		   
46168&	ML1\_20200510\_022851\_red&	$q$&	2020-05-10&	02:28:52.52&	yellow&19.75&	45.88 &	-    &-	           \\
46174&	ML1\_20200510\_024124\_red&	$q$&	2020-05-10&	02:41:25.25&	yellow&19.69&	48.16 &	19.15&	79.71	   \\
46180&	ML1\_20200510\_025352\_red&	$q$&	2020-05-10&	02:53:53.53&	yellow&19.59&	52.94 &	19.06&	86.17	   \\
46950&	ML1\_20200515\_025401\_red&	$q$&	2020-05-15&	02:54:01.10&	yellow&20.43&	24.35 &	19.41&	62.61	   \\
46957&	ML1\_20200515\_031152\_red&	$q$&	2020-05-15&	03:11:52.52&	yellow&20.33&	26.81 &	19.37&	64.83	   \\
46958&	ML1\_20200515\_031351\_red&	$q$&	2020-05-15&	03:13:52.52&	yellow&20.33&	26.74 &	19.37&	64.77	   \\
46959&	ML1\_20200515\_031552\_red&	$q$&	2020-05-15&	03:15:53.53&	yellow&20.31&	27.38 &	19.37&	65.08	   \\
46960&	ML1\_20200515\_031751\_red&	$q$&	2020-05-15&	03:17:51.51&	yellow&20.30&	27.55 &	19.35&	65.89	   \\
46961&	ML1\_20200515\_031950\_red&	$q$&	2020-05-15&	03:19:51.51&	yellow&20.22&	29.53 &	19.31&	68.44       \\
46962&	ML1\_20200515\_032150\_red&	$q$&	2020-05-15&	03:21:50.50&	yellow&20.41&	24.92 &	19.40&	63.12	    \\
46963&	ML1\_20200515\_032350\_red&	$q$&	2020-05-15&	03:23:51.51&	yellow&20.35&	26.23 &	19.38&	64.36	    \\
46964&	ML1\_20200515\_032549\_red&	$q$&	2020-05-15&	03:25:50.50&	yellow&20.14&	32.01 &	19.25&	72.18	    \\
46965&	ML1\_20200515\_032750\_red&	$q$&	2020-05-15&	03:27:51.51&	yellow&20.36&	25.99 &	19.38&	64.15	    \\
46966&	ML1\_20200515\_032951\_red&	$q$&	2020-05-15&	03:29:51.51&	yellow&20.31&	27.19 &	19.36&	65.17	    \\
46967&	ML1\_20200515\_033151\_red&	$q$&	2020-05-15&	03:31:52.52&	yellow&20.25&	28.78 &	19.34&	66.94	    \\
46968&	ML1\_20200515\_033351\_red&	$q$&	2020-05-15&	03:33:51.51&	yellow&20.38&	25.55 &	19.39&	63.47	    \\
46969&	ML1\_20200515\_033549\_red&	$q$&	2020-05-15&	03:35:49.49&	yellow&20.40&	25.11 &	19.40&	63.13  \\
46970&	ML1\_20200515\_033748\_red&	$q$&	2020-05-15&	03:37:48.48&	yellow&20.25&	28.94 &	19.33&	67.30  \\
46971&	ML1\_20200515\_033948\_red&	$q$&	2020-05-15&	03:39:48.48&	yellow&20.36&	25.95 &	19.39&	63.79  \\
46972&	ML1\_20200515\_034147\_red&	$q$&	2020-05-15&	03:41:48.48&	yellow&20.28&	27.94 &	19.37&	65.07  \\
46973&	ML1\_20200515\_034348\_red&	$q$&	2020-05-15&	03:43:48.48&	yellow&20.12&	32.45 &	19.28&	70.35  \\
46974&	ML1\_20200515\_034548\_red&	$q$&	2020-05-15&	03:45:48.48&	yellow&20.21&	30.02 &	19.32&	67.61  \\
46975&	ML1\_20200515\_034748\_red&	$q$&	2020-05-15&	03:47:49.49&	yellow&20.16&	31.21 &	19.32&	67.94  \\
46976&	ML1\_20200515\_034947\_red&	$q$&	2020-05-15&	03:49:47.47&	yellow&20.11&	32.83 &	19.29&	70.13  \\
46977&	ML1\_20200515\_035146\_red&	$q$&	2020-05-15&	03:51:47.47&	yellow&20.15&	31.61 &	19.31&	68.28  \\
46978&	ML1\_20200515\_035346\_red&	$q$&	2020-05-15&	03:53:47.47&	yellow&20.16&	31.20 &	19.32&	68.20  \\
46979&	ML1\_20200515\_035546\_red&	$q$&	2020-05-15&	03:55:46.46&	yellow&20.15&	31.67 &	19.31&	68.32  \\
46980&	ML1\_20200515\_035744\_red&	$q$&	2020-05-15&	03:57:45.45&	yellow&20.17&	30.96 &	19.32&	67.97  \\
46981&	ML1\_20200515\_035943\_red&	$q$&	2020-05-15&	03:59:44.44&	yellow&20.12&	32.55 &	19.30&	69.34  \\
46982&	ML1\_20200515\_040144\_red&	$q$&	2020-05-15&	04:01:44.44&	yellow&20.17&	31.12 &	19.32&	68.13  \\
46983&	ML1\_20200515\_040344\_red&	$q$&	2020-05-15&	04:03:44.44&	yellow&20.01&	35.83 &	19.22&	74.24  \\
46984&	ML1\_20200515\_040544\_red&	$q$&	2020-05-15&	04:05:45.45&	yellow&20.09&	33.30 &	19.28&	70.40  \\
46985&	ML1\_20200515\_040743\_red&	$q$&	2020-05-15&	04:07:43.43&	yellow&20.14&	31.98 &	19.31&	68.50  \\
46989&	ML1\_20200515\_041848\_red&	$q$&	2020-05-15&	04:18:49.49&	yellow&19.72&	47.16 &	19.13&	80.93  \\
46990&	ML1\_20200515\_042047\_red&	$q$&	2020-05-15&	04:20:48.48&	yellow&19.61&	52.06 &	19.07&	85.22  \\
46991&	ML1\_20200515\_042249\_red&	$q$&	2020-05-15&	04:22:49.49&	yellow&19.48&	58.84 &	19.00&	91.36  \\
46992&	ML1\_20200515\_042448\_red&	$q$&	2020-05-15&	04:24:49.49&	yellow&19.30&	69.12 &	18.89&	101.09 \\
46993&	ML1\_20200515\_042648\_red&	$q$&	2020-05-15&	04:26:48.48&	yellow&19.03&	88.90 &	18.67&	123.68 \\
\hline
\end{tabular}
\end{table*}

\newpage
\begin{table*}
\addtocounter{table}{-1}
\centering
	\caption{, continued}
	%\label{tab:meerlicht2}
	\begin{tabular}{rrrrrrrrrr} % four columns, alignment for each
		\hline
        Image ID & Image Name & Filter & Date & UT-mid & Flag & Lim. Mag & Lim. Flux & Lim Mag & Lim. Flux \\
		         & & & & & & Persistent & Persistent & Transient & Transient\\ 
		         & & & & & & (AB) & ($\mu$Jy) & (AB) & ($\mu$Jy)\\ \hline

47223&	ML1\_20200517\_025353\_red&	$q$&	2020-05-17&	02:53:53.53&	yellow&20.64&	20.15 &	19.44&	60.68  \\
47229&	ML1\_20200517\_031401\_red&	$q$&	2020-05-17&	03:14:02.20 &	yellow&20.56&	21.64 &	19.42&	61.73  \\
47230&	ML1\_20200517\_031600\_red&	$q$&	2020-05-17&	03:16:00.00 &	yellow&20.46&	23.68 &	19.39&	63.52  \\
47231&	ML1\_20200517\_031758\_red&	$q$&	2020-05-17&	03:17:58.58&	yellow&20.39&	25.42 &	19.34&	66.85  \\
47232&	ML1\_20200517\_031957\_red&	$q$&	2020-05-17&	03:19:57.57&	yellow&20.62&	20.50 &	19.44&	60.97  \\
47233&	ML1\_20200517\_032200\_red&	$q$&	2020-05-17&	03:22:00.00 &	yellow&20.64&	20.07 &	19.44&	61.03  \\
47234&	ML1\_20200517\_032400\_red&	$q$&	2020-05-17&	03:24:00.00 &	yellow&20.52&	22.48 &	19.40&	63.12  \\
47235&	ML1\_20200517\_032600\_red&	$q$&	2020-05-17&	03:26:01.10 &	yellow&20.63&	20.31 &	19.44&	60.95  \\
47236&	ML1\_20200517\_032800\_red&	$q$&	2020-05-17&	03:28:00.00 &	yellow&20.73&	18.60 &	19.45&	60.16  \\
47237&	ML1\_20200517\_033000\_red&	$q$&	2020-05-17&	03:30:01.10 &	yellow&20.52&	22.40 &	19.40&	62.90  \\
47238&	ML1\_20200517\_033623\_red&	$q$&	2020-05-17&	03:36:24.24&	red   &20.59&	21.03 &	-    &-	        \\
47239&	ML1\_20200517\_033823\_red&	$q$&	2020-05-17&	03:38:23.23&	yellow&20.39&	25.32 &	19.37&	64.88   \\
47240&	ML1\_20200517\_034022\_red&	$q$&	2020-05-17&	03:40:23.23&	yellow&20.42&	24.65 &	19.38&	64.24   \\
47241&	ML1\_20200517\_034221\_red&	$q$&	2020-05-17&	03:42:22.22&	yellow&20.60&	20.90 &	19.44&	61.05   \\
47242&	ML1\_20200517\_034422\_red&	$q$&	2020-05-17&	03:44:23.23&	yellow&20.56&	21.64 &	19.42&	61.76   \\
47243&	ML1\_20200517\_034623\_red&	$q$&	2020-05-17&	03:46:23.23&	yellow&20.41&	24.95 &	19.35&	65.91   \\
47244&	ML1\_20200517\_034822\_red&	$q$&	2020-05-17&	03:48:23.23&	yellow&20.63&	20.32 &	19.44&	60.81   \\
47245&	ML1\_20200517\_035022\_red&	$q$&	2020-05-17&	03:50:23.23&	yellow&20.51&	22.69 &	19.42&	61.72   \\
47246&	ML1\_20200517\_035220\_red&	$q$&	2020-05-17&	03:52:21.21&	red   &20.47&	23.56 &	-    &-	        \\
47247&	ML1\_20200517\_035422\_red&	$q$&	2020-05-17&	03:54:23.23&	yellow&20.59&	21.01 &	19.43&	61.18   \\
47248&	ML1\_20200517\_035622\_red&	$q$&	2020-05-17&	03:56:22.22&	yellow&20.57&	21.44 &	19.43&	61.19   \\
47249&	ML1\_20200517\_035822\_red&	$q$&	2020-05-17&	03:58:22.22&	yellow&20.41&	24.88 &	19.37&	65.03   \\
47250&	ML1\_20200517\_040022\_red&	$q$&	2020-05-17&	04:00:22.22&	yellow&20.57&	21.42 &	19.43&	61.29   \\
47251&	ML1\_20200517\_040222\_red&	$q$&	2020-05-17&	04:02:23.23&	yellow&20.62&	20.42 &	19.44&	60.97   \\
47252&	ML1\_20200517\_040422\_red&	$q$&	2020-05-17&	04:04:23.23&	yellow&20.52&	22.56 &	19.41&	62.41   \\
47253&	ML1\_20200517\_040621\_red&	$q$&	2020-05-17&	04:06:22.22&	yellow&20.42&	24.76 &	19.39&	63.78   \\
47254&	ML1\_20200517\_040821\_red&	$q$&	2020-05-17&	04:08:21.21&	yellow&20.46&	23.70 &	19.42&	62.10    \\
47258&	ML1\_20200517\_041930\_red&	$q$&	2020-05-17&	04:19:30.30&	yellow&19.99&	36.77 &	19.27&	71.29	 \\
47386&	ML1\_20200518\_025649\_red&	$q$&	2020-05-18&	02:56:50.50&	yellow&20.60&	20.82 &	19.42&	61.96	 \\
47390&	ML1\_20200518\_030502\_red&	$q$&	2020-05-18&	03:05:03.30 &	yellow&20.79&	17.59 &	19.46&	59.84	 \\
47391&	ML1\_20200518\_030701\_red&	$q$&	2020-05-18&	03:07:02.20 &	yellow&20.76&	18.03 &	19.45&	60.00	 \\
47392&	ML1\_20200518\_030901\_red&	$q$&	2020-05-18&	03:09:02.20 &	yellow&20.67&	19.59 &	19.44&	60.63	 \\
47393&	ML1\_20200518\_031100\_red&	$q$&	2020-05-18&	03:11:01.10 &	red   &20.70&	18.98 &	-    &-		 \\
47394&	ML1\_20200518\_031728\_red&	$q$&	2020-05-18&	03:17:29.29&	yellow&20.42&	24.71 &	19.41&	62.70	 \\
47395&	ML1\_20200518\_031927\_red&	$q$&	2020-05-18&	03:19:27.27&	yellow&20.33&	26.75 &	19.33&	67.15   \\
47396&	ML1\_20200518\_032126\_red&	$q$&	2020-05-18&	03:21:26.26&	yellow&20.64&	20.11 &	19.44&	60.62   \\
47397&	ML1\_20200518\_032324\_red&	$q$&	2020-05-18&	03:23:24.24&	yellow&20.53&	22.23 &	19.41&	62.75   \\
47398&	ML1\_20200518\_032522\_red&	$q$&	2020-05-18&	03:25:22.22&	yellow&20.52&	22.53 &	19.38&	64.19   \\
47399&	ML1\_20200518\_032721\_red&	$q$&	2020-05-18&	03:27:21.21&	yellow&20.76&	18.09 &	19.45&	60.07   \\
47400&	ML1\_20200518\_032921\_red&	$q$&	2020-05-18&	03:29:22.22&	yellow&20.74&	18.37 &	19.45&	60.34   \\
47401&	ML1\_20200518\_033119\_red&	$q$&	2020-05-18&	03:31:20.20&	yellow&20.56&	21.60 &	19.40&	62.96   \\
47402&	ML1\_20200518\_033319\_red&	$q$&	2020-05-18&	03:33:19.19&	yellow&20.78&	17.69 &	19.46&	59.78   \\
47403&	ML1\_20200518\_033517\_red&	$q$&	2020-05-18&	03:35:18.18&	yellow&20.77&	17.92 &	19.45&	60.02   \\
47404&	ML1\_20200518\_033715\_red&	$q$&	2020-05-18&	03:37:15.15&	yellow&20.49&	23.17 &	19.39&	63.68   \\
47405&	ML1\_20200518\_033914\_red&	$q$&	2020-05-18&	03:39:14.14&	yellow&20.66&	19.71 &	19.44&	60.83   \\
47406&	ML1\_20200518\_034113\_red&	$q$&	2020-05-18&	03:41:13.13&	yellow&20.79&	17.56 &	19.46&	59.92   \\
47407&	ML1\_20200518\_034311\_red&	$q$&	2020-05-18&	03:43:12.12&	yellow&20.64&	20.17 &	19.42&	62.08   \\
47408&	ML1\_20200518\_034512\_red&	$q$&	2020-05-18&	03:45:13.13&	yellow&20.76&	17.99 &	19.45&	60.19   \\
47409&	ML1\_20200518\_034712\_red&	$q$&	2020-05-18&	03:47:13.13&	yellow&20.64&	20.20 &	19.44&	60.69   \\
47410&	ML1\_20200518\_034913\_red&	$q$&	2020-05-18&	03:49:13.13&	red   &17.90&	252.03&	-    &-	        \\
47411&	ML1\_20200518\_035112\_red&	$q$&	2020-05-18&	03:51:13.13&	red   &18.85&	104.57&	-    &-	        \\
47412&	ML1\_20200518\_035310\_red&	$q$&	2020-05-18&	03:53:10.10&	yellow&20.70&	19.13 &	19.44&	60.91   \\
47413&	ML1\_20200518\_035512\_red&	$q$&	2020-05-18&	03:55:12.12&	yellow&20.62&	20.56 &	19.43&	61.30   \\
47417&	ML1\_20200518\_040618\_red&	$q$&	2020-05-18&	04:06:18.18&	yellow&20.52&	22.54 &	19.40&	63.13   \\
47418&	ML1\_20200518\_040816\_red&	$q$&	2020-05-18&	04:08:17.17&	yellow&20.56&	21.71 &	19.42&	61.80   \\
47419&	ML1\_20200518\_041017\_red&	$q$&	2020-05-18&	04:10:17.17&	yellow&20.56&	21.60 &	19.43&	61.54   \\
47420&	ML1\_20200518\_041217\_red&	$q$&	2020-05-18&	04:12:18.18&	yellow&20.53&	22.19 &	19.42&	62.20   \\
47421&	ML1\_20200518\_041416\_red&	$q$&	2020-05-18&	04:14:16.16&	yellow&20.54&	22.00 &	19.42&	61.83   \\
47422&	ML1\_20200518\_041615\_red&	$q$&	2020-05-18&	04:16:16.16&	yellow&20.31&	27.30 &	19.38&	64.44   \\
47423&	ML1\_20200518\_041814\_red&	$q$&	2020-05-18&	04:18:15.15&	yellow&20.34&	26.43 &	19.38&	64.20   \\
47424&	ML1\_20200518\_042013\_red&	$q$&	2020-05-18&	04:20:13.13&	yellow&20.23&	29.25 &	19.35&	65.80   \\
\hline
\end{tabular}
\end{table*}

\newpage
\begin{table*}
\addtocounter{table}{-1}
\centering
	\caption{, continued}
	%\label{tab:meerlicht3}
	\begin{tabular}{rrrrrrrrrr} % four columns, alignment for each
		\hline
        Image ID & Image Name & Filter & Date & UT-mid & Flag & Lim. Mag & Lim. Flux & Lim Mag & Lim. Flux \\
		         & & & & & & Persistent & Persistent & Transient & Transient\\ 
		         & & & & & & (AB) & ($\mu$Jy) & (AB) & ($\mu$Jy)\\ \hline

47471&	ML1\_20200519\_025439\_red&	$q$&	2020-05-19&	02:54:39.39&	red   &18.40&	159.06&	-    &-	        \\
47475&	ML1\_20200519\_030259\_red&	$q$&	2020-05-19&	03:02:59.59&	red   &20.11&	32.85 &	-    &-	        \\
47476&	ML1\_20200519\_030458\_red&	$q$&	2020-05-19&	03:04:58.58&	red   &20.29&	27.75 &	-    &-	        \\
47477&	ML1\_20200519\_030658\_red&	$q$&	2020-05-19&	03:06:59.59&	yellow&20.21&	29.80 &	19.41&	62.45   \\
47478&	ML1\_20200519\_030857\_red&	$q$&	2020-05-19&	03:08:58.58&	yellow&20.51&	22.66 &	19.45&	60.15   \\
47479&	ML1\_20200519\_031059\_red&	$q$&	2020-05-19&	03:10:59.59&	yellow&20.35&	26.33 &	19.30&	69.29   \\
47480&	ML1\_20200519\_031256\_red&	$q$&	2020-05-19&	03:12:57.57&	red   &19.81&	43.45 &	-    &-	         \\  
47481&	ML1\_20200519\_031544\_red&	$q$&	2020-05-19&	03:15:44.44&	red   &19.72&	47.15 &	-    &-		 \\
47482&	ML1\_20200519\_031742\_red&	$q$&	2020-05-19&	03:17:43.43&	red   &20.51&	22.66 &	-    &-		 \\
47483&	ML1\_20200519\_031941\_red&	$q$&	2020-05-19&	03:19:42.42&	yellow&20.46&	23.84 &	19.35&	65.84	 \\
47484&	ML1\_20200519\_032140\_red&	$q$&	2020-05-19&	03:21:40.40&	red   &20.31&	27.23 &	-    &-		 \\
47485&	ML1\_20200519\_032339\_red&	$q$&	2020-05-19&	03:23:40.40&	red   &20.67&	19.56 &	-    &-		 \\
47486&	ML1\_20200519\_032539\_red&	$q$&	2020-05-19&	03:25:40.40&	yellow&20.70&	19.14 &	19.28&	70.28	 \\
47487&	ML1\_20200519\_032740\_red&	$q$&	2020-05-19&	03:27:41.41&	yellow&20.16&	31.33 &	19.42&	62.12	 \\
47488&	ML1\_20200519\_032939\_red&	$q$&	2020-05-19&	03:29:40.40&	yellow&20.13&	32.19 &	19.25&	72.59    \\
47489&	ML1\_20200519\_033138\_red&	$q$&	2020-05-19&	03:31:38.38&	red   &19.83&	42.48 &	-    &-		 \\
47490&	ML1\_20200519\_033336\_red&	$q$&	2020-05-19&	03:33:37.37&	red   &19.52&	56.52 &	-    &-		 \\
47491&	ML1\_20200519\_033536\_red&	$q$&	2020-05-19&	03:35:36.36&	red   &19.05&	86.81 &	-    &-		 \\
47492&	ML1\_20200519\_033735\_red&	$q$&	2020-05-19&	03:37:35.35&	red   &19.90&	39.98 &	-    &-		 \\
47493&	ML1\_20200519\_033934\_red&	$q$&	2020-05-19&	03:39:34.34&	yellow&19.99&	36.76 &	19.41&	62.45	 \\
47494&	ML1\_20200519\_034132\_red&	$q$&	2020-05-19&	03:41:33.33&	red   &20.00&	36.27 &	-    &-		 \\
47495&	ML1\_20200519\_034332\_red&	$q$&	2020-05-19&	03:43:33.33&	red   &19.67&	49.33 &	-    &-		 \\
47496&	ML1\_20200519\_034532\_red&	$q$&	2020-05-19&	03:45:33.33&	red   &19.71&	47.27 &	-    &-	         \\
47497&	ML1\_20200519\_034735\_red&	$q$&	2020-05-19&	03:47:36.36&	red   &20.07&	34.01 &	-    &-		 \\
47498&	ML1\_20200519\_034933\_red&	$q$&	2020-05-19&	03:49:33.33&	yellow&20.43&	24.37 &	19.45&	60.15	 \\
47501&	ML1\_20200519\_041930\_red&	$q$&	2020-05-19&	04:19:30.30&	red   &15.06&	3449.61&-     &	-	 \\
48107&	ML1\_20200521\_031954\_red&	$q$&	2020-05-21&	03:19:54.54&	yellow&20.16&	31.25 &	19.30&	69.29	 \\
     &                         &         &                &                &          &     &         &      &          \\  		    	             	       	      		 
46169&	ML1\_20200510\_023055\_red&	$r$&	2020-05-10&	02:30:56.56&	yellow&18.95&	95.33 &	-    &	-	 \\
46175&	ML1\_20200510\_024328\_red&	$r$&	2020-05-10&	02:43:29.29&	yellow&19.08&	84.99 &	18.22&	187.69	 \\
46181&	ML1\_20200510\_025557\_red&	$r$&	2020-05-10&	02:55:58.58&	yellow&19.19&	76.63 &	18.29&	175.99   \\
46951&	ML1\_20200515\_025606\_red&	$r$&	2020-05-15&	02:56:06.60 &	yellow&19.72&	46.85 &	18.43&	154.13	 \\
47224&	ML1\_20200517\_025556\_red&	$r$&	2020-05-17&	02:55:56.56&	yellow&19.78&	44.34 &	18.41&	157.72	 \\
47472&	ML1\_20200519\_025643\_red&	$r$&	2020-05-19&	02:56:43.43&	red   &18.66&	124.89&	18.46&	149.83	 \\
     &                         &         &                &                &          &     &         &      &           \\ 			        	  		   		    	             	       	      		  
46170&	ML1\_20200510\_023300\_red&	$i$&	2020-05-10&	02:33:01.10 &	yellow&18.79&	110.50&	-    &	-          \\
46176&	ML1\_20200510\_024532\_red&	$i$&	2020-05-10&	02:45:33.33&	yellow&18.79&	111.03&	18.22&	187.69	  \\
46182&	ML1\_20200510\_025802\_red&	$i$&	2020-05-10&	02:58:03.30 &	yellow&18.91&	99.28 &	18.29&	175.99	  \\
46952&	ML1\_20200515\_025811\_red&	$i$&	2020-05-15&	02:58:11.11&	yellow&19.32&	67.76 &	18.43&	154.13	  \\
47225&	ML1\_20200517\_025802\_red&	$i$&	2020-05-17&	02:58:02.20 &	yellow&19.25&	72.75 &	18.41&	157.72	  \\
47388&	ML1\_20200518\_030049\_red&	$i$&	2020-05-18&	03:00:50.50&	yellow&19.49&	58.23 &	18.46&	149.83	  \\
47473&	ML1\_20200519\_025848\_red&	$i$&	2020-05-19&	02:58:48.48&	red   &18.06&	216.36&	-    &	-	  \\
     &                         &         &                &                &          &     &         &      &           \\ 			        	  		   		    	             	       	      		  
46171&	ML1\_20200510\_023505\_red&	$z$&	2020-05-10&	02:35:06.60 &	yellow&17.92&	246.81&	-    &	-	 \\
46177&	ML1\_20200510\_024738\_red&	$z$&	2020-05-10&	02:47:39.39&	yellow&17.86&	259.62&	17.30&	436.24	 \\
46183&	ML1\_20200510\_030007\_red&	$z$&	2020-05-10&	03:00:08.80 &	yellow&17.98&	233.71&	17.37&	409.28   \\
46953&	ML1\_20200515\_030018\_red&	$z$&	2020-05-15&	03:00:18.18&	yellow&18.20&	191.12&	17.46&	376.41	 \\
47226&	ML1\_20200517\_030008\_red&	$z$&	2020-05-17&	03:00:08.80 &	yellow&18.04&	220.28&	17.38&	406.98	 \\
47389&	ML1\_20200518\_030256\_red&	$z$&	2020-05-18&	03:02:57.57&	yellow&18.17&	196.21&	17.44&	385.04	 \\
47474&	ML1\_20200519\_030054\_red&	$z$&	2020-05-19&03:00:54.54&	red   &17.56&	342.28&	-    &	-	 \\  
\hline
     \end{tabular}
\end{table*}

%%%%%%%%%%%%%%%%%%%%%%%%%%%%%%%%%%%%%%%%%%%%%%%%%%

% Don't change these lines
\bsp	% typesetting comment
\label{lastpage}
\end{document}